\pdfoutput=1
\documentclass[aps,prb,
twocolumn,
superscriptaddress,nobibnotes]{revtex4-1}

\usepackage{amsmath}
\usepackage{graphicx}
\usepackage{multirow}
\usepackage{color}
\usepackage{bm}

\usepackage[normalem]{ulem}

\definecolor{gray}{gray}{0.5}
\usepackage[colorlinks=true,linkcolor=blue,citecolor=blue,urlcolor=black]{hyperref}
\pdfpageattr{/Group << /S /Transparency /I true /CS /DeviceRGB>>}
\begin{document}

\title{Strain-induced polar discontinuities in two-dimensional materials from combined first-principles and Schr\"odinger--Poisson simulations}

\author{Augustin Bussy}
\author{Giovanni Pizzi}
\author{Marco Gibertini}
\affiliation{Theory and Simulation of Materials (THEOS) and National Centre for Computational Design and Discovery of Novel Materials (MARVEL), \'Ecole Polytechnique F\'ed\'erale de Lausanne, CH-1015 Lausanne, Switzerland}

\begin{abstract}
The local application of mechanical stress in piezoelectric materials gives rise to boundaries across which the electric polarization changes. Polarization charges appear along such polar discontinuities and the ensuing electric fields drive a charge reconstruction with the accumulation of free carriers at the boundaries.
This is particularly relevant for two-dimensional materials that can sustain very large strains and display record piezoelectric responses. Here we show by first-principles simulations the emergence of one-dimensional wires of free electrons and holes along strain interfaces, taking SnSe as a paradigmatic material. We complement this by developing a Schr\"odinger--Poisson approach specifically designed for two-dimensional materials that it is able to reproduce the ab-initio results and also to extend them to regimes of parameters and system sizes that would be unaffordable in first principles calculations. This model allows us to assess the degree of tunability for the free charge in the wires coming from strain values and profiles, and to obtain the critical size at which the interfaces start to be metallic. 
\end{abstract}

\maketitle
\section{Introduction}

A polar discontinuity is a discrete change in electric polarization across an interface or surface. 
Classical electrostatics tells us that bound polarization charges appear at such polar discontinuities and the system  responds with compensating charges in order to screen the corresponding electric fields. 
In many polar surfaces and interfaces this happens through atomic reconstructions and changes in stoichiometry~\cite{Harrison1978,Noguera2000,Goniakowski2008}. Alternatively, a charge reconstruction can occur, with the accumulation of free electrons or holes at the polar discontinuity to screen the polarization charges. 
This mechanism is considered to be at the origin of the two-dimensional (2D) electron gas appearing at the polar interface between insulating perovskites like LaAlO$_3$ and SrTiO$_3$~\cite{Oht2004,Bris2014}.

Recently, the emergence of metallic states at polar discontinuities has been extended to 2D materials~\cite{MartinezGordillo2015}. In this case, one-dimensional (1D) wires of free carriers have been theoretically predicted to appear at polar interfaces between different 2D materials~\cite{Bristowe2013,Gibertini2014}, at boundaries between functionalized and pristine monolayers~\cite{Gibertini2014},  along the edges of polar nanoribbons~
\cite{Gul2013,Gibertini2014,Gibertini2015}, and at inversion domain boundaries~\cite{Gibertini2015}. Possible applications include electronics, spintronics and solar-energy harvesting.
Experimental signatures of such metallic states have been reported in nanostructures based on transition metal dichalcogenides, such as triangular islands of MoS$_2$~\cite{Bol2001} and twin boundaries in MoSe$_2$~\cite{Liu2014,Barja2016}. In the latter case, the rational value of the amount of charge per unit length in the wires (2/3) makes the system susceptible to Peierls instabilities~\cite{Peierls}, giving rise to charge-density waves with the opening of a small gap at the Fermi energy~\cite{Barja2016}. Although being of utmost fundamental interest, charge density waves might hinder potential applications of these systems. 

In this respect, piezoelectricity might offer new strategies to induce polar discontinuities that are more resilient to charge-density waves.
In fact, the application of mechanical stress on piezoelectric materials induces a change in their electric polarization. 
When stress is applied only locally, a boundary appears between strained and unstrained regions, i.e., between regions with different electric polarizations. This means that polar discontinuities occur at strain interfaces in piezoelectric materials.  
This is particularly relevant for 2D materials that have been shown to withstand very large strains~\cite{Lee385,Akinwande2017} and are endowed with sizable piezoelectric responses~\cite{Duerloo2012,fei2015giant,Gomes2015,Blonsky2015}. 
In particular, group-IV monochalcogenides like SnSe and SnTe exhibit colossal piezoelectric coefficients~\cite{fei2015giant,Gomes2015} and a ferroelectric polarization that can be switched with an external electric field~\cite{Fei2016,Hanakata2016,Chang2016}. 
Here we show by accurate first-principles simulations that interfaces between strained and unstrained SnSe host 1D wires of free carriers as a consequence of the polar discontinuity associated with the piezoelectric response of the material. 
The amount of free electrons and holes in these wires can be controlled by the 	value of strain and it is in general not commensurate with the unit length, thus making these metallic states more robust against Peierls instabilities. 

In the context of  3D semiconductor heterostructures the self-consistent solution of coupled Schr\"odinger and Poisson equations has turned out to be extremely useful to predict the charge distribution in doped multilayer systems and even in the case of polar discontinuities in undoped 3D oxides~\cite{Janotti2012}.  In order to gain additional insight into the properties of strain interfaces in SnSe and complement first-principles results, here we also develop a Schr\"odinger--Poisson (SP) solver specifically designed for 2D materials that properly takes into account the unconventional non-local screening occurring in 2D~\cite{Keldysh1979,cudazzo2011dielectric,Qiu2016}.
The solution of the 2D SP equations has been validated against first-principles results for the charge density and spatial distribution of free carriers at strain interfaces in SnSe. 
The much cheaper SP approach allows us to extend ab initio calculations and investigate a wide range of strain values and systems sizes that would be otherwise unaffordable.
This allows us to assess the tunability of the amount of free charge in the wires by strain and to obtain the critical size at which the interfaces start to be metallic. 
Finally we show that the assumption of sharp strain interfaces is not necessary and that the effect is robust against a gradual  change in strain across an interface.

\section{Methods}

\subsection{First-principles simulations}
First-principles simulations are performed within density-functional theory (DFT) using the Quantum ESPRESSO~\cite{Gia2009} suite of codes. The Perdew--Burke--Ernzerhof~\cite{PBE} generalized-gradient approximation of the exchange-correlation functional is adopted. Electron-ion interactions are accounted for using ultrasoft pseudopotentials from the GBRV library~\cite{GBRV} that have been tested~\cite{SSSP} to obtain the best accuracy with respect to all-electron calculations for the elements considered in this work (Sn and Se). An energy cutoff of 40 (320) Ry is employed to expand wave-functions (densities) in all cases, with the exception of variable-cell calculations for which the cutoffs were doubled.  The Brillouin zone is sampled using a $14\times14\times1$ Monkhorst--Pack grid~\cite{Monk1976} for bulk 2D materials (both strained and unstrained) while  for strain interfaces we used a $1\times20\times1$ grid with a 0.005 Ry Marzari-Vanderbilt smearing~\cite{marzari1999thermal}. Bulk structures were carefully relaxed with a threshold of 10$^{-3}$~eV/\AA\ for forces and 0.5 kbar for stresses using a Broyden--Fletcher--Goldfarb--Shanno algorithm. 2D systems are simulated within 3D periodic boundary conditions by using the effective-screening-medium approach~\cite{Otani2006} with 20~\AA\ of vacuum to eliminate the spurious effect of artificial periodic replicas. 	

\subsection{2D Schr\"odinger--Poisson Solver}
In the macroscopic limit we can assume 2D materials to have vanishing thickness and, for definiteness, to lie in the $xy$ plane. We are interested in the case when the system is homogeneous and infinitely extended along the $y$ direction. 
The electrostatic potential $\phi$ is then a function of $x$ and $z$ only. The Poisson equation links $\phi(x,z)$ to the total charge density $\rho_{\rm tot}(x,z)$, which is the sum of a free $\rho_{\rm f}$ and polarization $\rho_{\rm pol}$ contribution. In this framework, both densities are confined to the plane, \textit{i.e.} $\rho_{\rm f/pol}(x,z) = \sigma_{\rm f/pol}(x)\delta(z)$, with $\sigma_{\rm f/pol}(x)$ being the planar density in the 2D sheet. The polarization charge density depends on the polarizability $\alpha$ of the material and, generalizing the expression by Cudazzo \textit{et al.}~\cite{cudazzo2011dielectric} to a spatially-varying $\alpha$, we obtain $\sigma_{\rm pol}(x)=\partial_x\left[ \alpha(x)\partial_x\phi(x,z=0) \right]$. The relevant Poisson equation is hence:
\begin{align}\label{poisson1}
\nabla^2\phi(x,z) & =  -4\pi \sigma_{\rm tot}(x)\delta(z),
\end{align}
where
\begin{align}\label{sigmatot}
\sigma_{\rm tot} & =  \sigma_{\rm f}(x) + \partial_x\left[ \alpha(x)\partial_x\phi(x,z=0) \right].
\end{align}
Given the total charge density, the Poisson equation~\eqref{poisson1} can be solved using the superposition principle:
\begin{equation}\label{unscreened_pois}
\phi(x,z) =  \int \text{d}x' \sigma_{\rm tot}(x') \phi_{\rm w}(x'-x,z), \\
\end{equation}
where $\phi_{\rm w}(x,z) $ is the potential produced by an infinite wire along $y$ with unit linear charge density. In open boundary conditions we have $\phi^{\rm OBC}_{\rm w}(x,z)  = -\log(x^2+z^2)$, while in periodic boundary conditions $\phi^{\rm PBC}_{\rm w}(x,z)  = -\log[2\cosh(2\pi z/L)- 2\cos(2\pi x/L)]$, $L$ being the periodicity along $x$ (see Supplemental Material~\cite{suppl}). Since the right-hand side  of Eq.~\eqref{unscreened_pois}  depends on the in-plane electrostatic potential, this equation can be solved self-consistently with Eq.~\eqref{sigmatot}; we choose the free charge density as a starting point.

We use this solution of the Poisson equation to implement a multi-band SP solver suited for 2D materials. Since we assume homogeneity along $y$, the Schr\"odinger equation is one-dimensional and easy to solve numerically using finite-difference techniques~\cite{brown1990numerical,tan1990self}. We work at zero temperature, in the parabolic-band approximation and use Pulay mixing~\cite{pulay1980convergence} as convergence scheme. 
In addition to the peculiar nature of screening in 2D described above, another key difference with respect to 3D SP solvers such as in Refs.~\cite{serra1991one,karner2007multi,Busby2010} is that the density of states (DOS) of a one-dimensional electron gas is used to fill the bands. Finally, we introduce some cold smearing~\cite{marzari1999thermal} of the zero temperature Fermi--Dirac distribution for improved convergence and for fair comparison with first-principles calculations. 
The zero-temperature assumption is required to compare our solver with the DFT results. This assumption can be easily lifted in our solver. The correct materials parameters (like effective masses and position of the band edges), however, should be used, as these change with temperature due to effects like electron-phonon interaction.
The code implementing this solver is released open-source on GitHub~\cite{schrpoisson}.

\begin{figure*}
\begin{center}
\includegraphics{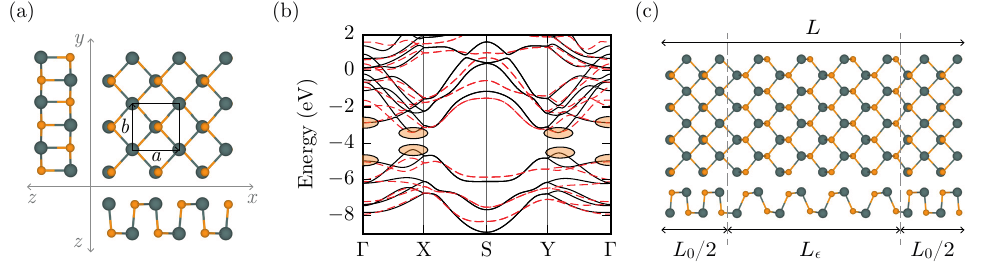}
\end{center}
\caption{Panel (a):  top and lateral views of monolayer SnSe. Panel (b): band structure along a high-symmetry path in the Brillouin zone of pristine SnSe (black solid lines) and upon uniaxial straining with $\epsilon=0.1$ along the $x$ direction (red dashed lines). Valleys of interest for SP simulations are located at $\Gamma$, along $\Gamma$-X and along $\Gamma$-Y, both for valence and conduction bands. The strained data was rescaled to fit the plot. Panel (c): top and lateral views of interfaces between strained (central) and unstrained (sides) monolayer SnSe. The strain is uniaxial and along the $x$ direction.  We assume to have the same number  $N$ of strained and unstrained unit cells (here $N=4$) and the supercell size $L$ along $x$ is is the sum of the widths $L_\epsilon$ and $L_0$ of the strained and unstrained regions, respectively.
\label{fig:structure}}
\end{figure*}

The parameters needed for the SP solver are obtained via DFT calculations on  bulk 2D materials. The effective mass tensors are extracted by computing the Hessian matrix of energy bands close to their extrema.  In the case studied in the following, effective masses are diagonal in the Cartesian coordinate system, with the  $xx$ component representing  the confinement mass and the $yy$ component being the DOS mass. Energy bands are aligned by referring them to the vacuum level, defined as the electrostatic potential far away from the 2D layer.  Finally, the polarizability $\alpha$ can be computed using density-functional perturbation theory (DFPT) from the scaling of the in-plane dielectric constant in a repeated-layer model with respect to the interlayer separation~\cite{cudazzo2011dielectric,Sohier2016}.

\section*{Results and discussion}

We consider SnSe as a paradigmatic monolayer to illustrate the emergence of polar discontinuities and 1D wires of free electrons/holes across strain interfaces in 2D materials. 
Owing to its large piezoelectric response~\cite{fei2015giant,Gomes2015}, SnSe offers the best platform to engineer large polar discontinuities at moderate strain and it is thus the optimal candidate material to validate our predictions in experiments. 

In Fig.~\ref{fig:structure}(a) we show top and lateral views of the crystal structure typical of group-IV monochalcogenides. In the case of SnSe, the rectangular unit cell has relaxed lattice parameters $a=4.41$~\AA\ and $b=4.29$~\AA\ along the $x$ and $y$ directions, respectively. A finite intrinsic electric polarization ${\bm P}_0$ along  $x$ is present as a consequence of the horizontal displacement between Sn and Se atoms. The electromechanical coupling in SnSe allows to tune the polarization value by straining the material (piezoelectric effect)~\cite{fei2015giant,Gomes2015}, while its direction can be switched by applying an in-plane electric field (ferroelectric effect)~\cite{Fei2016,Hanakata2016}, which has been recently validated experimentally in a related material, SnTe~\cite{Chang2016}. 
The band structure of monolayer SnSe is reported in Fig.~\ref{fig:structure}(b) (black solid lines), showing a finite gap with the valence band maximum along the $\Gamma-$X direction and two almost degenerate conduction band minima along the $\Gamma-$X and  $\Gamma-$Y directions, respectively. The zero of energy is set at the vacuum level, that is at the electrostatic potential far away from the monolayer. In the same panel we also plot the band structure when a uniaxial strain $\epsilon=0.1$ is applied along the $x$ direction (red dashed lines), that is parallel to the intrinsic polarization. The energy location of conduction band minima is only marginally affected by strain, while a larger effect is manifest on effective masses (see also Supplemental Material) and on the valence band maximum, giving rise to an increase of band gap. As a consequence of its piezoelectric response, such strained SnSe will also display a different electric polarization. This means that at interfaces between strained and unstrained SnSe, as the ones shown in Fig.~\ref{fig:structure}(c), a polar discontinuity will occur. The linear density of polarization charges accumulating at such strain interfaces increases with the amount of strain (see Supplemental Material) and is given by 
\begin{equation}\label{eq:lambdaP}
\lambda_{\rm P} = - \hat{\bm n}\cdot({\bm P}_0 - {\bm P}_{\epsilon}),
\end{equation}
 $\hat{\bm n}$ being the unit vector normal to the interface and pointing from strained to unstrained regions. 
The polarization charge density $\lambda_{\rm P}$ crucially depends on the interface orientation and it is maximal in modulus when the interface is along a direction perpendicular to the change in polarization (as in Fig.~\ref{fig:structure}c and in all simulations hereafter).  

 \begin{figure*} 
\begin{center}
\includegraphics{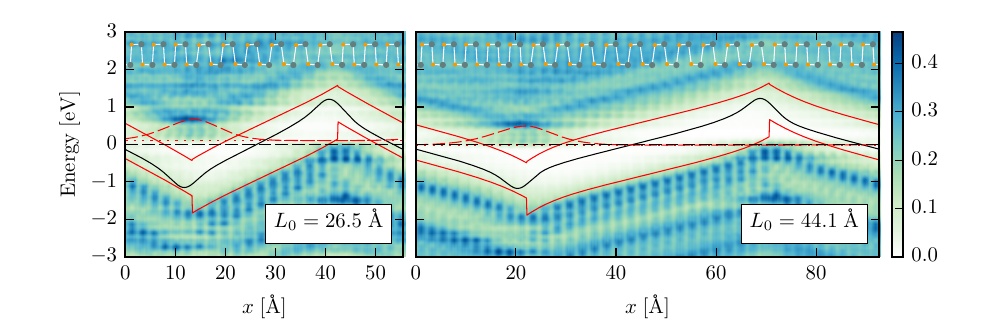}
\end{center}
\caption{Color plot of the local density of states as a function of energy and position across strain interfaces with $L_0=26.5~{\rm \AA}$ (corresponding to $N=6$, left) and $L_0=44.1~{\rm \AA}$ (corresponding to $N=10$, right) and $\epsilon=0.1$, computed with DFT. The zero of energy is set at the Fermi energy. The solid line is the 2D effective electrostatic potential discussed in the text, while the red solid lines are the band profiles according to Schr\"odinger--Poisson simulations, showing an excellent match with the first-principles simulations. The spatial profile of the lowest state arising from conduction bands is reported with a red dashed line, while the dotted red line represents its energy position.}
\label{fig:ldos}
\end{figure*}

In the following we consider only uniaxial strain $\epsilon$ along the polarization direction $x$, so that the strained unit cell size along $x$ is $a_\epsilon =(1+ \epsilon)\, a$, while both strained and unstrained cells have the same length $b$ along $y$.
In order to simulate strain interfaces from first-principles within periodic boundary conditions, we take alternating stripes of strained and unstrained SnSe obtained by periodically repeating the structure in Fig.~\ref{fig:structure}(c) along the $x$ direction. We assume to have the same number  $N$ of strained and unstrained unit cells, so that the supercell size along $x$ is $L=L_\epsilon+L_0$, with $L_\epsilon = N a_\epsilon = (1+\epsilon) N a= (1+\epsilon) L_0$. 
To complement first-principles calculations we take advantage of our SP solver and simulate the system by  modelling it as a sequential repetition of  strained and unstrained regions of lateral size $L_\epsilon$ and $L_0$, respectively, separated by sharp interfaces across which effective masses, band edges, and polarizabilities change abruptly. Three valence and three conduction bands, with their respective degeneracies, are taken into account and are associated with the local maxima/minima at the $\Gamma$ point and along the $\Gamma-$X and $\Gamma-$Y directions emphasized in Fig.~\ref{fig:structure}(b). Band extrema are referred to the vacuum level, resulting in larger offsets in the valence bands than in conduction bands (see also Supplemental Figure 3). Polarization charges, computed from DFT using a Berry phase approach~\cite{King1992,Resta2007}, are included as $\delta$-doping at the interfaces in the SP solver~\cite{Janotti2012}. Unless otherwise stated, we use  a discretization step  $\Delta x = 0.2$~\AA\ to solve the Schr\"odinger and Poisson equations. 

In Fig.~\ref{fig:ldos} we plot the local density-of-states (LDOS) computed with DFT integrated over the vertical $z$ coordinate and averaged over $y$, as a function of energy and position along $x$ for two systems with $L_0=26.5~$\AA\ and $44.1~$\AA, corresponding respectively to $N=6$ and $N=10$. Locally a band gap is present between valence and conduction states, but the band edges become position dependent. In order to understand this behaviour we also plot the in-plane macroscopic electrostatic potential (black solid line) that can be obtained from Eq.~\eqref{unscreened_pois} at $z=0$ by taking as $\sigma_{\rm tot}(x)$ the macroscopic average~\cite{Bal1988} of the total charge density integrated over $z$ and averaged over $y$. The presence of polarization charges is clearly evident from the non-vanishing electric field appearing throughout the system. Away from the interfaces, the electric field gives rise to a simple shift of the energy states, so that the band edges follow the electrostatic potential. Close to the interfaces, more subtle quantum effects take place as a consequence of confinement  associated with the approximately logarithmic potential generated by the polarization charges. To clarify this, we performed SP simulations for the same systems. In Fig.~\ref{fig:ldos} red solid lines represent the SP band edge profiles and show a remarkable agreement with the DFT potential and LDOS away from the interfaces. We also report the energy (red dash-dotted line) and profile (red dashed line) of the lowest-energy state arising from conduction bands in SP simulations. This state is localized at the left interface and quantum confinement endows it with an energy that lies significantly above the SP conduction band profile. This is in very good agreement with the LDOS from DFT and explains the flat band edge profile in the LDOS close to the interfaces and its deviation from the electrostatic potential. We have thus elucidated the importance of quantum effects in the emergence of free carriers at polar discontinuities in 2D materials, contrary to their 3D counterparts where semiclassical arguments can be still applied to explain the relationship between electrostatic potentials and band profiles~\cite{Bris2014}. 

\begin{figure} 
\begin{center}
\hspace{1.5cm}\includegraphics[width=0.8\linewidth]{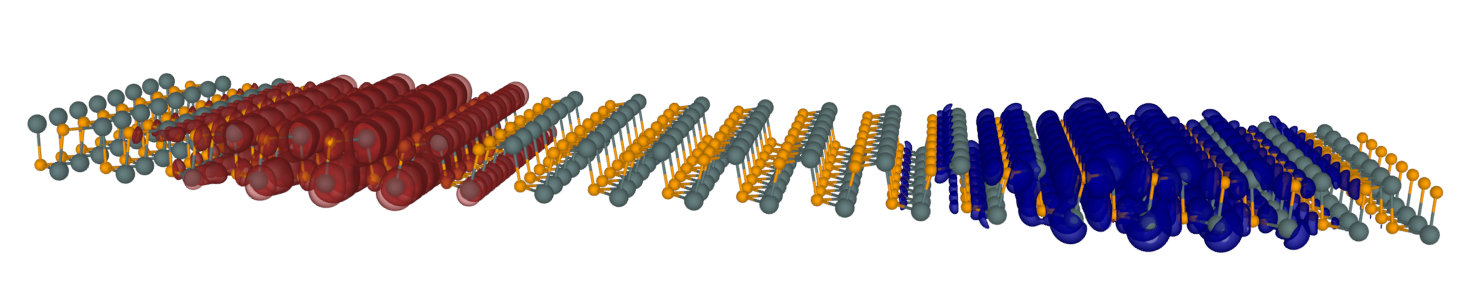}\\
\includegraphics{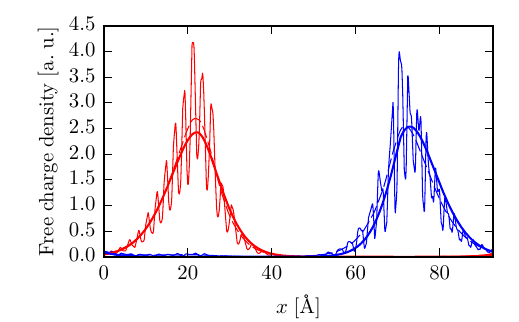}
\end{center}
\caption{Top panel: 3D spatial distribution of the free electron (red) and hole (blue) density from DFT forming 1D wires along the interface between strained and unstrained monolayer SnSe (for a system with $L_0=44.1~{\rm \AA}$, i.e. $N = 10$, and $\epsilon = 0.1$). 
Bottom panel: linear charge density of free electrons (red) and holes (blue) as a function of the coordinate $x$ across the interface for the same system as in the top panel. Thin solid and dashed lines are the bare and macroscopically-averaged results from DFT. Thick solid lines are obtained from SP simulations.
\label{fig:dens_prof}}
\end{figure}

It is also interesting to follow the evolution of this lowest-energy conduction state as a function of the system size. For $L_0=26.5~$\AA\ ($N=6$) this state is empty and the system is semiconducting, although the gap is considerably reduced as a consequence of the electric fields that tend to shift upwards the valence states on one interface and downwards the conduction states on the other. For $L_0=44.1~$\AA\ ($N=10$) the system is metallic as this state lies below the Fermi energy and gets partially occupied. This signals the occurrence of an insulator-to-metal transition with increasing $N$. The partial occupation of this conduction state and the corresponding partial depletion of a valence state lead to the formation of pockets of free electrons and free holes. These pockets are localized at opposite interfaces and form 1D metallic wires of free electrons and holes that extend along the interfaces and partially screen the polarization charges. This is even more clear in Fig.~\ref{fig:dens_prof} where we show the 3D spatial profile of the free carrier density obtained by integrating the LDOS separately for electrons (red) and holes (blue) when $L_0=44.1~$\AA\ and $\epsilon=0.1$. Averaging these 3D free-carrier densities over $y$ and integrating over $z$ gives the profiles reported in the main panel (thin lines). From SP simulations we obtain density profiles (thick lines) that are in very good agreement with the macroscopic average of the DFT results (dashed lines), validating once more the predictiveness of the 2D SP solver.  

\begin{figure} 
\begin{center}
\includegraphics{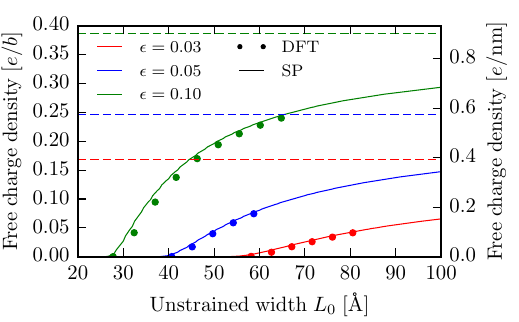}
\end{center}
\caption{Linear density of free electrons/holes as a function of the unstrained region width $L_0$ for three different values of strain ($\epsilon=0.03$, 0.05, and 0.1). Circles correspond to the DFT charge density obtained by integrating the LDOS, while solid lines to SP data. Dashed lines correspond to the asymptotic values of the free-charge density for $L_0\to\infty$ given by $|\lambda_{\rm P}|$ in Eq.~\eqref{eq:lambdaP}.\label{fig:dens_vs_strain}}
\end{figure}

The linear density of free electrons/holes present in the 1D wires depends on the system size, as shown in Fig.~\ref{fig:dens_vs_strain}. DFT results (symbols) are reported for three different values of strain ($\epsilon = 0.03$, 0.05, and 0.1) and are in remarkable agreement with the same quantity obtained from SP simulations (solid lines), highlighting the predictive quality of the SP solver. An insulator-to-metal transition occurs for any value of strain as the system size is increased. The critical width of the unstrained region at which free carriers start to populate the wires decreases with increasing strain, passing from $\approx 57$~\AA\ for $\epsilon=0.03$ to $\approx 26$~\AA\ for $\epsilon=0.1$. This can be understood from the fact that with increasing strain the change in electric polarization across the interfaces is enhanced, together with the modulus of the polarization charge density in Eq.~\eqref{eq:lambdaP} (see also Supplemental Material). As a consequence, the electric fields are larger and the overall potential drop between the interfaces increases with strain. This favours the transition to a metallic state, that is obtained at smaller critical widths when the localized conduction state on one interface and the hole state on the opposite one meet at the Fermi energy. By increasing the width beyond the critical value, the amount of free carriers accumulated in the wires increases, until asymptotically the free charge density perfectly balances the polarization charge density (dashed lines in Fig.~\ref{fig:dens_vs_strain}) and the overall charge at each interface vanishes.

\begin{figure} 
\begin{center}
\includegraphics{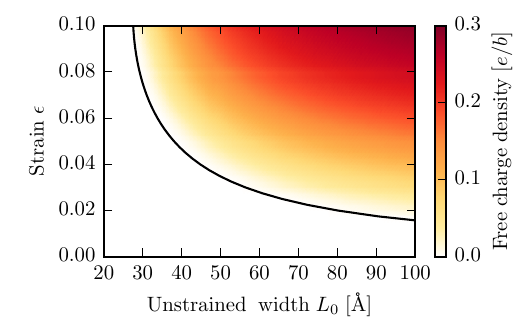}
\end{center}
\caption{Free-carrier density as a function of strain $\epsilon$ and width of the unstrained region $L_0$ obtained with the 2D Schr\"odinger--Poisson solver. The black solid line marks the prediction for the insulator-to-metal transition. \label{fig:predictions}}
\end{figure}

The excellent agreement between DFT and SP simulations allows us to rely on the much less expensive SP approach to compute the free carrier density over a wide range of system sizes and strain values at the same predictive level of DFT calculations. To reduce even further the computational cost, effective masses, band edges, polarizabilities and all necessary input parameters for SP simulations at  arbitrary strain  are obtained by fitting the corresponding DFT results over a finite set of strain values with low-order polynomials (see Supplemental Material). 
Results for the linear charge density of free electrons/holes as a function of strain and lateral size of the unstrained regions are reported in Fig.~\ref{fig:predictions}. 
As expected from the discussion above, the free-carrier density increases with increasing strain and with increasing system size. In particular, this allows us to obtain the critical width of the unstrained region beyond which the system becomes metallic as a function of strain $\epsilon$ in the strained region (black solid line). This critical width decreases with increasing strain, with a minimal value of $\sim2.6$~nm for the largest value of $\epsilon$ considered here. 

\begin{figure} 
\begin{center}
\includegraphics{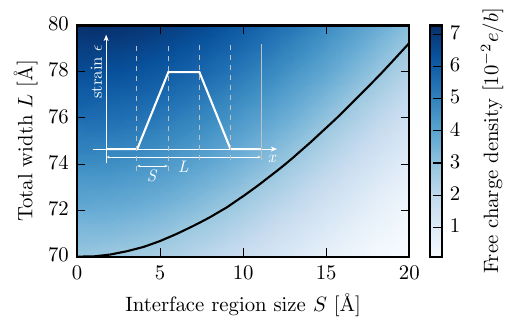}
\end{center}
\caption{\label{fig:sharpinterfaces}Free-carrier density as a function of the total width of the system $L$  and of the length of the interface region $S$, for a maximum strain $\epsilon=0.08$ in the central region. The black line is a contour of constant free-charge density corresponding to $L=70$~\AA\ and sharp interfaces ($S=0$). Inset: Strain profile used in the simulation.}
\end{figure}

Up to now, we have always considered systems with sharp strain interfaces. This choice was mainly dictated by the need to restrain system sizes within the limits of DFT simulations and to simplify the description of the physical interpretation of the results. However, in a realistic setup the strain profile would be a continuous function of position. We want to show now that the phenomena reported above do not require sharp interfaces. Instead, the occurrence of polar discontinuities (and the magnitude of the charge density in the channels) is largely independent of the interface details.
To prove this claim, we perform simulations of a setup having the strain profile illustrated in the inset of Fig.~\ref{fig:sharpinterfaces} using our SP solver, that is capable to consider any smooth strain profile. In particular, we consider a total system width $L$ in the range 70--80~\AA, with the strain profile changing from a minimum of zero to a maximum of 8\% over a range of length $S$. For illustration purposes, we interpolate linearly the strain between the two constant-strain regions, but we verified that the results would be only marginally affected by a change in the strain profile. The lengths of each of the two constant-strain regions are kept equal to $(L-2S)/2$ in each simulation, and we choose values of S from 0 to 20~\AA\ (that is, up to interface regions very large compared to the system size, with $2S/L$ up to 50\%).
Although 2D materials can give rise to finite flexoelectric responses~\cite{Yudin2013,Naumov2009}, we do not include them in simulations as they would only affect the profile of polarization charge density and not the overall amount of free carriers in the wires (especially in the asymptotic limit of large system sizes).
In Fig.~\ref{fig:sharpinterfaces} we show a colour plot of the free-carrier density as a function of the geometrical parameters $L$ and $S$. We also show a contour line at constant free-carrier density at the value obtained for $L=70$~\AA\ and $S=0$~\AA, i.e. with sharp interfaces. The contour line makes it immediately evident that even for $S>0$ it is possible to obtain the same charge density that would be obtained for sharp interfaces (apart from the need of a slight adaptation of the total system length $L$, which is expected as the effective separation between the wires is reduced for increasing values of $S$).
In the Supplemental Material we also show a comparison of the band edge profile and of the first confined states for two different values of $S$, showing that apart from a smoothing of the band profile, the electronic properties of the system are very similar.
These results show the feasibility of the approach that we present here to induce polar discontinuities in 2D materials also in realistic situations where the strain might change gradually over several unit cells rather than abruptly.

\section*{Conclusions}

We have shown that the local application of strain in piezoelectric 2D materials can give rise to polar discontinuities across the interfaces between strained and unstrained regions. Such interfaces could be realized by suspending the sample over an expanding gap or depositing it over neighbouring substrates with different expansion coefficients~\cite{Duerloo2014}. By using density-functional theory simulations for the prototypical material SnSe, we have demonstrated the occurrence of an insulator-to-metal transition with the appearance of 1D wires of free electrons and holes along the interface. To further characterize the system, we have developed a Schr\"odinger--Poisson approach specifically designed for 2D materials that takes into account the subtleties associated with screening in low dimensions and that has been validated against first-principles results. The code implementing this solver is released open-source on GitHub~\cite{schrpoisson} and allowed us to compute the free-carrier density over a wide range of system sizes and strain values, and thus to assess the critical width at which the insulator-to-metal transition occurs as a function of strain and the effect of smooth strain profiles. 

\section*{Acknowledgements}
We would like to thank Nicola Marzari for useful discussions and support. 
This project has received partial funding from the Graphene Flagship through the European Union's Horizon 2020 research and innovation programme under grant agreement No 696656 GrapheneCore1.
We also acknowledge simulation time from the Swiss National Supercomputing Centre (CSCS) through project ID s580.

\onecolumngrid

\renewcommand\figurename{Supplementary Fig.}
\renewcommand\tablename{Supplementary Table}
\renewcommand\theequation{S\arabic{equation}}
\setcounter{equation}{0}
\setcounter{figure}{0}
\setcounter{table}{0}

\newpage
\section*{Supplementary Data}

\subsection*{Potential generated by a periodic repetition of charged wires}

We are interested in the potential generated by a periodic repetition of charged wires with linear charge density $\lambda$, separated by a distance $L$ along the $x$ direction and extending infinitely along $y$. Although more sophisticated derivations based on complex-plane representations of the problem, we prefer to adopt a more intuitive approach starting from the superposition principle. 
Indeed, the potential $\phi^{\rm PBC}_{\rm w}(x,z)$ can be simply computed as the sum of the potentials generated by each wire:
\begin{align}
\phi^{\rm PBC}_{\rm w}(x,z) &  =  \sum_{n=-\infty}^{+\infty} \left\{-\lambda\ln\left[(x-nL)^2+z^2\right]\right\} = -\lambda\ln \prod_{n=-\infty}^{+\infty}\left[(x-nL)^2+z^2\right]\notag\\
&=-\lambda\ln\left\{ (x^2+z^2) \prod_{n=1}^{+\infty}\left[(x-nL)^2+z^2\right]\left[(x+nL)^2+z^2\right] \right\}\notag\\
&=-\lambda\ln\left\{ (x^2+z^2) \prod_{n=1}^{+\infty}(x- n L + i z) (x-nL-i z) (x +  n L + i z) (x+nL-i z)\right\}\notag\\
&=-\lambda\ln\left\{ (x^2+z^2) \prod_{n=1}^{+\infty}\left[n^2 L^2 - (x + i z)^2\right] \left[n^2L^2-(x-i z)^2\right] \right\} \notag\\
&=-\lambda\ln\left\{ (x+iz)(x-iz) \prod_{n=1}^{+\infty} n^4L^4\left[1- \frac{(x + i z)^2}{n^2L^2}\right] \left[1-\frac{(x-i z)^2}{n^2L^2}\right] \right\}\notag\\
&=-\lambda\ln\left\{ (x+iz) \prod_{n=1}^{+\infty} \left[1- \frac{(x + i z)^2}{n^2L^2}\right] (x-iz) \prod_{n=1}^{+\infty}\left[1-\frac{(x-i z)^2}{n^2L^2}\right] \right\} + K\notag\\
&=-\lambda\ln\left[4\sin\left(\pi\frac{x+iz}{L}\right)  \sin\left(\pi\frac{x-iz}{L}\right)\right] + K+c =-\lambda\ln\left[4\left|\sin\left(\pi\frac{x+iz}{L}\right)  \right|^2\right] + K+c  \notag\\
&=-2\lambda\ln\left|2\sin\left(\pi\frac{x+iz}{L}\right)  \right| + K+c  \notag\\
&=-\lambda\ln\left[2\cosh(2\pi z/L)-2\cos(2\pi x/L)  \right] + K+c
\end{align}
Here is $K$ is an infinite constant that we set to zero by requiring that the potential on a wire generated by all other wires is finite and we  made use of Euler's formula $\sin(\zeta)=\zeta\prod_{n=1}^{\infty}\left(1-\frac{\zeta^2}{\pi^2n^2}\right)$. The second constant $c$ can  also be set to zero in order to have that the potential far away from the wires is identical to the the one generated by a charged plane with planar density $\lambda/L$, i.e. $\phi^{\rm PBC}_{\rm w}(x,z\to\infty) = -2\pi \lambda/L |z|$. We thus finally have 
\begin{equation}
\boxed{\phi^{\rm PBC}_{\rm w}(x,z)  = -2\lambda\ln\left|2\sin\left(\pi\frac{x+iz}{L}\right)  \right| =-\lambda\ln\left[2\cosh(2\pi z/L)-2\cos(2\pi x/L)  \right] }
\end{equation}

\begin{table} [h]
\begin{center}
\begin{tabular*}{\textwidth}{l @{\extracolsep{\fill}} c c c c c }
\hline\hline \\[-2ex]
 & \multicolumn{2}{c}{Valence Band} & &\multicolumn{2}{c}{Conduction Band} \\
\hline \\[-1ex]
 & Unstrained & 10\% & & Unstrained & 10\% \\[0.1ex]
 \hline \\[-1ex]
 
 $m_{xx}^{\Gamma}$ & 1.756  & 0.804 & & 2.741 & 0.985 \\[2ex]
 $m_{yy}^{\Gamma}$ & 2.733 & 2.748 & & 2.994  & 1.593 \\[2ex]
 $m_{xx}^{\rm \Gamma-X}$ & 0.125 & 0.206 & & 0.111 & 0.183 \\[2ex]
 $m_{yy}^{\rm \Gamma-X}$ & 0.159 & 1.060 & & 0.190 & 0.213 \\[2ex]
 $m_{xx}^{\rm \Gamma-Y}$ & 0.110 & 0.271 & & 0.132 & 0.206 \\[2ex]
 $m_{yy}^{\rm \Gamma-Y}$ & 0.160 & 0.689 & & 0.130 & 0.320 \\[2ex]

\hline\hline \\[-2ex]
\end{tabular*}
\caption{Effective masses  in units of electron  mass obtained by diagonalizing the Hessian of the DFT band structure of bulk SnSe close to a minimum/maximum. Data are reported for $\epsilon = 0.1$ strain and for the unstrained material. The $xx$ subscript corresponds to the confinement mass used in the Schr\"odinger equation and $yy$ to the mass used in density of states (DOS) calculations.  }
\end{center}
\end{table}

\begin{figure} [h]
\begin{center}
\centerline{\includegraphics[width=10cm]{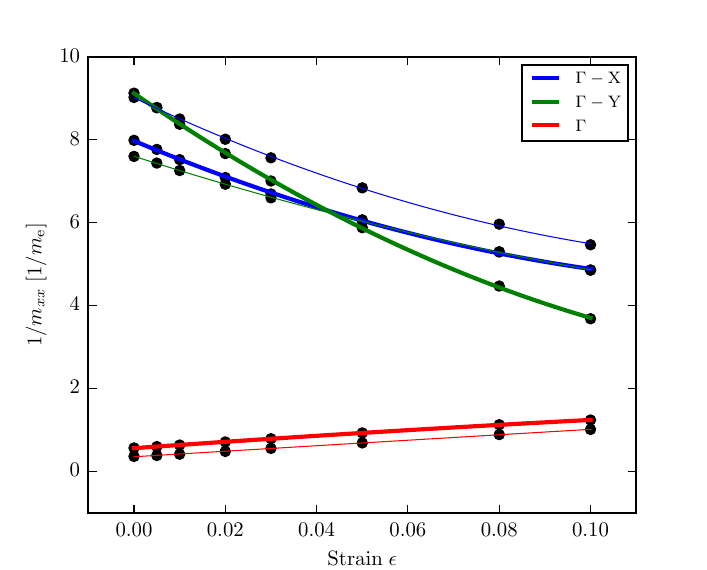}}
\caption{Quadratic fitting of the inverse confinement effective masses from DFT. Thick lines correspond to the valence band and thin lines to the conduction band.}
\end{center}
\end{figure}

\begin{figure} [h]
\begin{center}
\centerline{\includegraphics[width=10cm]{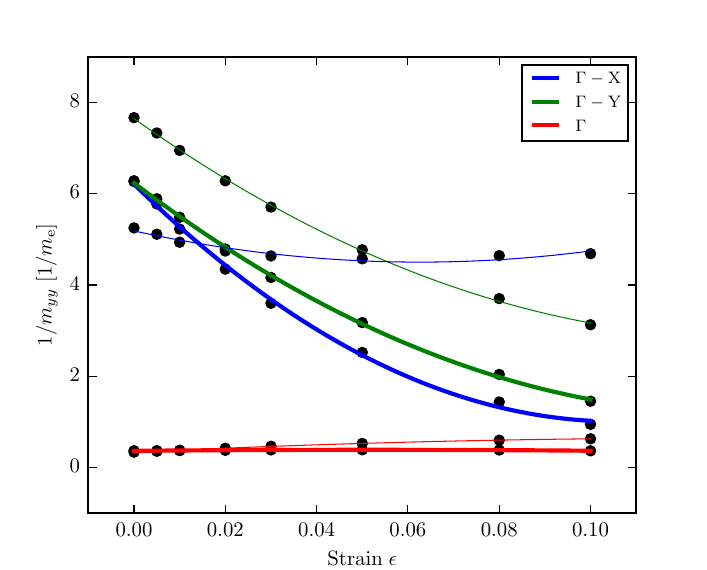}}
\caption{Quadratic fitting of the inverse DOS effective masses from DFT. Thick lines correspond to the valence band and thin lines to the conduction band.}
\end{center}
\end{figure}

\begin{figure} [h]
\begin{center}
\centerline{\includegraphics[width=10cm]{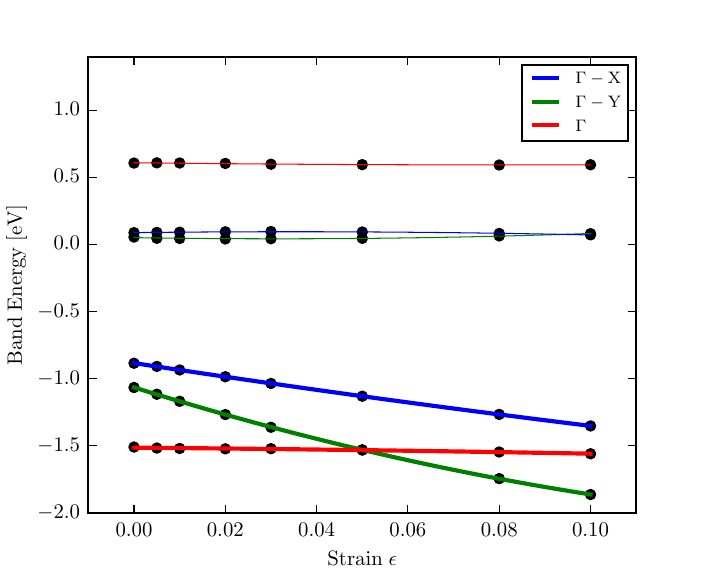}}
\caption{Quadratic fitting of bulk SnSe band structures extrema. Thick lines correspond to the valence band and thin lines to the conduction band. Values at $\epsilon = 0$ are $E_v^{\Gamma} = -1.508$ eV, $E_v^{\rm \Gamma-X} = -0.883$ eV and $E_v^{\rm \Gamma-Y} = -1.064$ eV for the valence band and $E_c^{\Gamma} = 0.609$ eV, $E_c^{\rm \Gamma-X} = 0.090$ eV and $E_c^{\rm \Gamma-Y} = 0.057$ eV for the conduction band.}
\end{center}
\end{figure}

\begin{figure} [h]
\begin{center}
\centerline{\includegraphics[width=10cm]{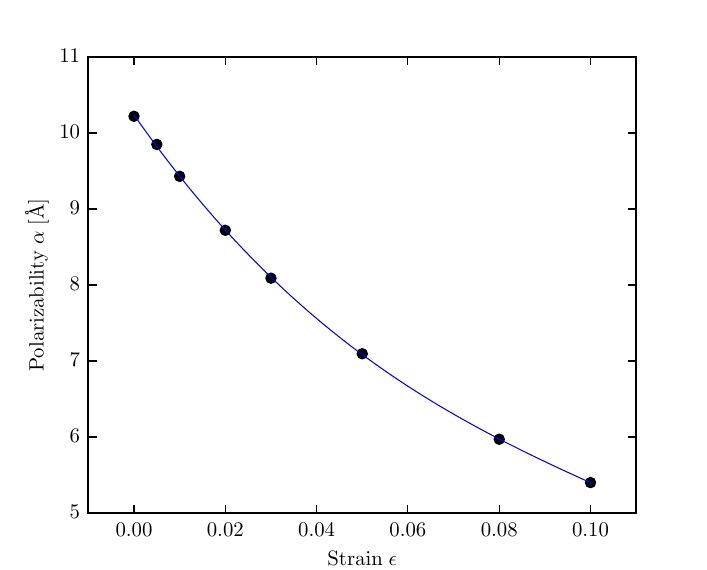}}
\caption{Cubic fitting of the polarizability $\alpha$ of monolayer SnSe in the $x$ direction. The value at zero strain is $\alpha_{\epsilon = 0} = 10.22 \AA$.}
\end{center}
\end{figure}

\begin{figure} [h]
\begin{center}
\centerline{\includegraphics[width=10cm]{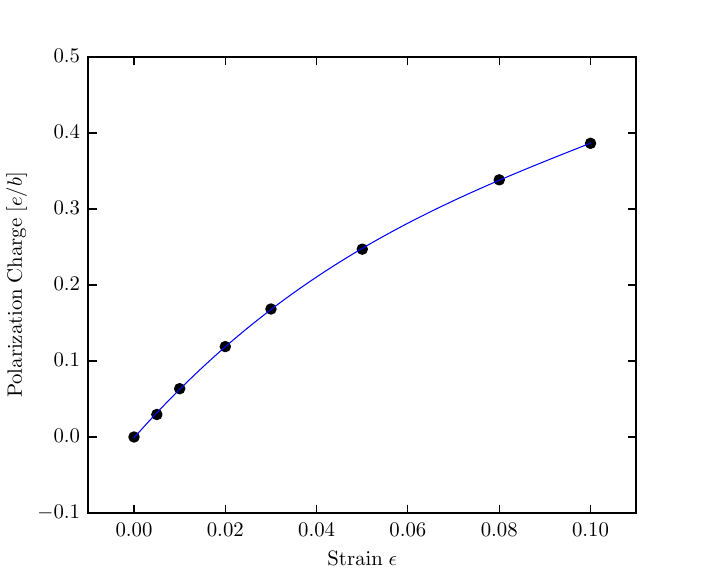}}
\caption{Cubic fitting of the polarization charge density at interfaces between unstrained and uniaxially strained monolayer SnSe with strain $\epsilon$. The unit of the polarization charge density is $e/b$, where $b$ is the lattice vector in the unstrained $y$ direction. }
\end{center}
\end{figure}

\begin{figure} 
\begin{center}
\includegraphics[width=10cm]{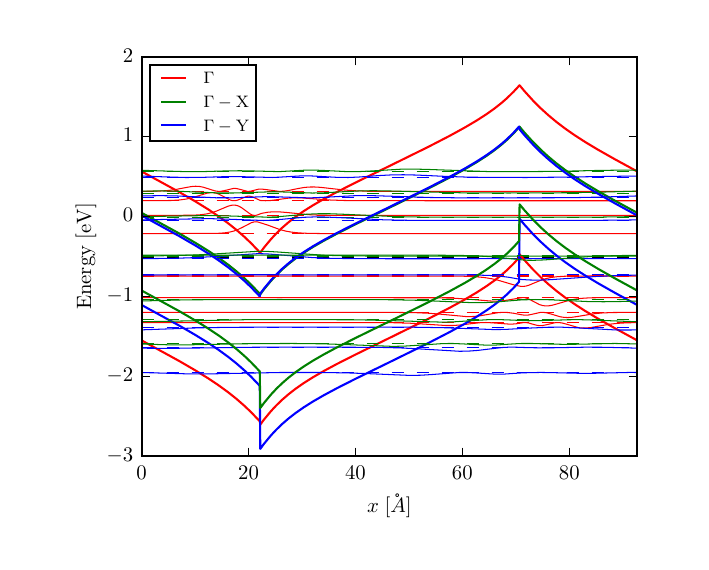}
\end{center}
\caption{Band profile of a setup with unstrained  width of $L_0 = 44.08$ \AA\ corresponding to $N = 10$ cells and with strain $\epsilon = 0.1$. Thick lines correspond to the energy of the conduction/valence bands minima/maxima, thin lines to the square modulus of the lowest-energy quantum states wavefunctions and thin dashed lines to their energy. The thick black dashed line is the Fermi level. 6 bands are considered with extrema at $\Gamma$, along $\Gamma$-X (doubly degenerate) and $\Gamma$-Y (doubly degenerate).}
\label{fig:band_prof}
\end{figure}

\begin{figure} 
\begin{center}
\includegraphics[width=8cm]{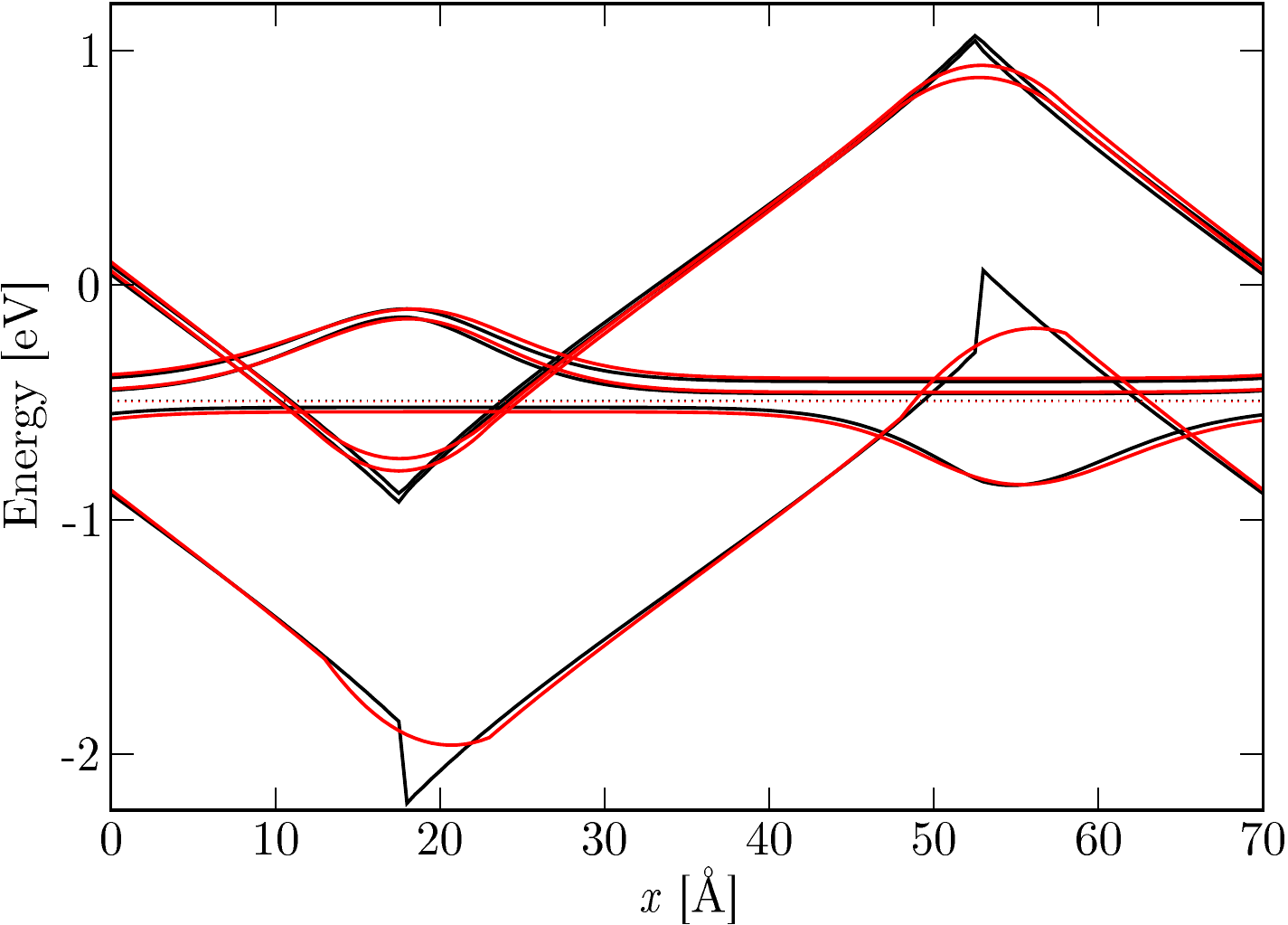}
\end{center}
\caption{Comparison of the relevant band profiles and electronic states of two setups with total length $L=70$~\AA\, with sharp interfaces ($S=0$~\AA, black) and with an interface region of length $S=10$~\AA\ (red). The definition of $S$ is given in the inset of Fig.~6 in the main text. The maximum strain in the central region is $\epsilon = 0.08$ in both cases. The dotted lines are the Fermi levels for the two systems. We consider only the relevant band edges (two in the conduction and one in the valence bands). For each edge, we only show the occupied states (only one per edge). Apart from a smearing of the band edge profile near the interfaces, no major change occurs in the electronic structure when the interface is not sharp ($S>0$).}
\label{fig:edge_comparison}
\end{figure}


\begin{thebibliography}{10}%
\makeatletter
\providecommand \@ifxundefined [1]{%
 \ifx #1\undefined \expandafter \@firstoftwo
 \else \expandafter \@secondoftwo
\fi
}%
\providecommand \@ifnum [1]{%
 \ifnum #1\expandafter \@firstoftwo
 \else \expandafter \@secondoftwo
\fi
}%
\providecommand \enquote [1]{``#1''}%
\providecommand \bibnamefont  [1]{#1}%
\providecommand \bibfnamefont [1]{#1}%
\providecommand \citenamefont [1]{#1}%
\providecommand\href[0]{\@sanitize\@href}%
\providecommand\@href[1]{\endgroup\@@startlink{#1}\endgroup\@@href}%
\providecommand\@@href[1]{#1\@@endlink}%
\providecommand \@sanitize [0]{\begingroup\catcode`\&12\catcode`\#12\relax}%
\@ifxundefined \pdfoutput {\@firstoftwo}{%
 \@ifnum{\z@=\pdfoutput}{\@firstoftwo}{\@secondoftwo}%
}{%
 \providecommand\@@startlink[1]{\leavevmode}%
 \providecommand\@@endlink[0]{}%
}{%
 \providecommand\@@startlink[1]{%
  \leavevmode
  \pdfstartlink
   attr{/Border[0 0 1 ]/H/I/C[0 1 1]}%
   user{/Subtype/Link/A<</Type/Action/S/URI/URI(#1)>>}%
  \relax
 }%
 \providecommand\@@endlink[0]{\pdfendlink}%
}%
\providecommand \url  [0]{\begingroup\@sanitize \@url }%
\providecommand \@url [1]{\endgroup\@href {#1}{\urlprefix}}%
\providecommand \urlprefix [0]{URL }%
\providecommand \Eprint[0]{\href }%
\@ifxundefined \urlstyle {%
  \providecommand \doi [1]{doi:\discretionary{}{}{}#1}%
}{%
  \providecommand \doi [0]{doi:\discretionary{}{}{}\begingroup
  \urlstyle{rm}\Url }%
}%
\providecommand \doibase [0]{http://dx.doi.org/}%
\providecommand \Doi[1]{\href{\doibase#1}}%
\providecommand \bibAnnote [3]{%
  \BibitemShut{#1}%
  \begin{quotation}\noindent
    \textsc{Key:}\ #2\\\textsc{Annotation:}\ #3%
  \end{quotation}%
}%
\providecommand \bibAnnoteFile [2]{%
  \IfFileExists{#2}{\bibAnnote {#1} {#2} {\input{#2}}}{}%
}%
\providecommand \typeout [0]{\immediate \write \m@ne }%
\providecommand \selectlanguage [0]{\@gobble}%
\providecommand \bibinfo [0]{\@secondoftwo}%
\providecommand \bibfield [0]{\@secondoftwo}%
\providecommand \translation [1]{[#1]}%
\providecommand \BibitemOpen[0]{}%
\providecommand \bibitemStop [0]{}%
\providecommand \bibitemNoStop [0]{.\EOS\space}%
\providecommand \EOS [0]{\spacefactor3000\relax}%
\providecommand \BibitemShut [1]{\csname bibitem#1\endcsname}%
\bibitem{Harrison1978}%
  \BibitemOpen
  \bibfield{author}{%
  \bibinfo {author} {\bibfnamefont{W.~A.}\ \bibnamefont{Harrison}}, \bibinfo
  {author} {\bibfnamefont{E.~A.}\ \bibnamefont{Kraut}}, \bibinfo {author}
  {\bibfnamefont{J.~R.}\ \bibnamefont{Waldrop}},\ and\ \bibinfo {author}
  {\bibfnamefont{R.~W.}\ \bibnamefont{Grant}},\ }%
  \Doi{10.1103/PhysRevB.18.4402}{\emph{\bibinfo {title} {Polar heterojunction
  interfaces}}},\ \bibfield{journal}{%
  \bibinfo {journal} {Phys. Rev. B}\ }%
  \textbf{\bibinfo {volume} {18}},\ \bibinfo {pages} {4402} (\bibinfo {year}
  {1978}).%
  \bibAnnoteFile{Stop}{Harrison1978}%
\bibitem{Noguera2000}%
  \BibitemOpen
  \bibfield{author}{%
  \bibinfo {author} {\bibfnamefont{C.}~\bibnamefont{Noguera}},\ }%
  \href{http://stacks.iop.org/0953-8984/12/i=31/a=201}{\emph{\bibinfo {title}
  {Polar oxide surfaces}}},\ \bibfield{journal}{%
  \bibinfo {journal} {Journal of Physics: Condensed Matter}\ }%
  \textbf{\bibinfo {volume} {12}},\ \bibinfo {pages} {R367} (\bibinfo {year}
  {2000}).%
  \bibAnnoteFile{Stop}{Noguera2000}%
\bibitem{Goniakowski2008}%
  \BibitemOpen
  \bibfield{author}{%
  \bibinfo {author} {\bibfnamefont{J.}~\bibnamefont{Goniakowski}}, \bibinfo
  {author} {\bibfnamefont{F.}~\bibnamefont{Finocchi}},\ and\ \bibinfo {author}
  {\bibfnamefont{C.}~\bibnamefont{Noguera}},\ }%
  \href{http://stacks.iop.org/0034-4885/71/i=1/a=016501}{\emph{\bibinfo {title}
  {Polarity of oxide surfaces and nanostructures}}},\ \bibfield{journal}{%
  \bibinfo {journal} {Reports on Progress in Physics}\ }%
  \textbf{\bibinfo {volume} {71}},\ \bibinfo {pages} {016501} (\bibinfo {year}
  {2008}).%
  \bibAnnoteFile{Stop}{Goniakowski2008}%
\bibitem{Oht2004}%
  \BibitemOpen
  \bibfield{author}{%
  \bibinfo {author} {\bibfnamefont{A.}~\bibnamefont{Ohtomo}}\ and\ \bibinfo
  {author} {\bibfnamefont{H.~Y.}\ \bibnamefont{Hwang}},\ }%
  \Doi{10.1038/nature02308}{\emph{\bibinfo {title} {A high-mobility electron
  gas at the LaAlO3/SrTiO3 heterointerface}}},\ \bibfield{journal}{%
  \bibinfo {journal} {Nature}\ }%
  \textbf{\bibinfo {volume} {427}},\ \bibinfo {pages} {423} (\bibinfo {year}
  {2004}).%
  \bibAnnoteFile{Stop}{Oht2004}%
\bibitem{Bris2014}%
  \BibitemOpen
  \bibfield{author}{%
  \bibinfo {author} {\bibfnamefont{N.~C.}\ \bibnamefont{Bristowe}}, \bibinfo
  {author} {\bibfnamefont{P.}~\bibnamefont{Ghosez}}, \bibinfo {author}
  {\bibfnamefont{P.~B.}\ \bibnamefont{Littlewood}},\ and\ \bibinfo {author}
  {\bibfnamefont{E.}~\bibnamefont{Artacho}},\ }%
  \Doi{10.1088/0953-8984/26/14/143201}{\emph{\bibinfo {title} {The origin of
  two-dimensional electron gases at oxide interfaces: insights from theory}}},\
  \bibfield{journal}{%
  \bibinfo {journal} {Journal of Physics: Condensed Matter}\ }%
  \textbf{\bibinfo {volume} {26}},\ \bibinfo {pages} {143201} (\bibinfo {year}
  {2014}).%
  \bibAnnoteFile{Stop}{Bris2014}%
\bibitem{MartinezGordillo2015}%
  \BibitemOpen
  \bibfield{author}{%
  \bibinfo {author} {\bibfnamefont{R.}~\bibnamefont{Martinez-Gordillo}}\ and\
  \bibinfo {author} {\bibfnamefont{M.}~\bibnamefont{Pruneda}},\ }%
  \Doi{http://dx.doi.org/10.1016/j.progsurf.2015.08.001}{\emph{\bibinfo {title}
  {Polar discontinuities and 1D interfaces in monolayered materials}}},\
  \bibfield{journal}{%
  \bibinfo {journal} {Progress in Surface Science}\ }%
  \textbf{\bibinfo {volume} {90}},\ \bibinfo {pages} {444 } (\bibinfo {year}
  {2015}).%
  \bibAnnoteFile{Stop}{MartinezGordillo2015}%
\bibitem{Bristowe2013}%
  \BibitemOpen
  \bibfield{author}{%
  \bibinfo {author} {\bibfnamefont{N.~C.}\ \bibnamefont{Bristowe}}, \bibinfo
  {author} {\bibfnamefont{M.}~\bibnamefont{Stengel}}, \bibinfo {author}
  {\bibfnamefont{P.~B.}\ \bibnamefont{Littlewood}}, \bibinfo {author}
  {\bibfnamefont{E.}~\bibnamefont{Artacho}},\ and\ \bibinfo {author}
  {\bibfnamefont{J.~M.}\ \bibnamefont{Pruneda}},\ }%
  \Doi{10.1103/PhysRevB.88.161411}{\emph{\bibinfo {title} {One-dimensional
  half-metallic interfaces of two-dimensional honeycomb insulators}}},\
  \bibfield{journal}{%
  \bibinfo {journal} {Phys. Rev. B}\ }%
  \textbf{\bibinfo {volume} {88}},\ \bibinfo {pages} {161411} (\bibinfo {year}
  {2013}).%
  \bibAnnoteFile{Stop}{Bristowe2013}%
\bibitem{Gibertini2014}%
  \BibitemOpen
  \bibfield{author}{%
  \bibinfo {author} {\bibfnamefont{M.}~\bibnamefont{Gibertini}}, \bibinfo
  {author} {\bibfnamefont{G.}~\bibnamefont{Pizzi}},\ and\ \bibinfo {author}
  {\bibfnamefont{N.}~\bibnamefont{Marzari}},\ }%
  \href{http://dx.doi.org/10.1038/ncomms6157}{\emph{\bibinfo {title}
  {Engineering polar discontinuities in honeycomb lattices}}},\
  \bibfield{journal}{%
  \bibinfo {journal} {Nat. Commun.}\ }%
  \textbf{\bibinfo {volume} {5}},\ \bibinfo {pages} {5157} (\bibinfo {year}
  {2014}).%
  \bibAnnoteFile{Stop}{Gibertini2014}%
\bibitem{Gul2013}%
  \BibitemOpen
  \bibfield{author}{%
  \bibinfo {author} {\bibfnamefont{F.}~\bibnamefont{G\"uller}}, \bibinfo
  {author} {\bibfnamefont{A.~M.}\ \bibnamefont{Llois}}, \bibinfo {author}
  {\bibfnamefont{J.}~\bibnamefont{Goniakowski}},\ and\ \bibinfo {author}
  {\bibfnamefont{C.}~\bibnamefont{Noguera}},\ }%
  \Doi{10.1103/PhysRevB.87.205423}{\emph{\bibinfo {title} {Polarity effects in
  unsupported polar nanoribbons}}},\ \bibfield{journal}{%
  \bibinfo {journal} {Phys. Rev. B}\ }%
  \textbf{\bibinfo {volume} {87}},\ \bibinfo {pages} {205423} (\bibinfo {year}
  {2013}).%
  \bibAnnoteFile{Stop}{Gul2013}%
\bibitem{Gibertini2015}%
  \BibitemOpen
  \bibfield{author}{%
  \bibinfo {author} {\bibfnamefont{M.}~\bibnamefont{Gibertini}}\ and\ \bibinfo
  {author} {\bibfnamefont{N.}~\bibnamefont{Marzari}},\ }%
  \Doi{10.1021/acs.nanolett.5b02834}{\emph{\bibinfo {title} {Emergence of
  One-Dimensional Wires of Free Carriers in Transition-Metal-Dichalcogenide
  Nanostructures}}},\ \bibfield{journal}{%
  \bibinfo {journal} {Nano Letters}\ }%
  \textbf{\bibinfo {volume} {15}},\ \bibinfo {pages} {6229} (\bibinfo {year}
  {2015}).%
  \bibAnnoteFile{Stop}{Gibertini2015}%
\bibitem{Bol2001}%
  \BibitemOpen
  \bibfield{author}{%
  \bibinfo {author} {\bibfnamefont{M.~V.}\ \bibnamefont{Bollinger}}, \bibinfo
  {author} {\bibfnamefont{J.~V.}\ \bibnamefont{Lauritsen}}, \bibinfo {author}
  {\bibfnamefont{K.~W.}\ \bibnamefont{Jacobsen}}, \bibinfo {author}
  {\bibfnamefont{J.~K.}\ \bibnamefont{N\o{}rskov}}, \bibinfo {author}
  {\bibfnamefont{S.}~\bibnamefont{Helveg}},\ and\ \bibinfo {author}
  {\bibfnamefont{F.}~\bibnamefont{Besenbacher}},\ }%
  \Doi{10.1103/PhysRevLett.87.196803}{\emph{\bibinfo {title} {One-Dimensional
  Metallic Edge States in ${\mathrm{MoS}}_{2}$}}},\ \bibfield{journal}{%
  \bibinfo {journal} {Phys. Rev. Lett.}\ }%
  \textbf{\bibinfo {volume} {87}},\ \bibinfo {pages} {196803} (\bibinfo {year}
  {2001}).%
  \bibAnnoteFile{Stop}{Bol2001}%
\bibitem{Liu2014}%
  \BibitemOpen
  \bibfield{author}{%
  \bibinfo {author} {\bibfnamefont{H.}~\bibnamefont{Liu}}, \bibinfo {author}
  {\bibfnamefont{L.}~\bibnamefont{Jiao}}, \bibinfo {author}
  {\bibfnamefont{F.}~\bibnamefont{Yang}}, \bibinfo {author}
  {\bibfnamefont{Y.}~\bibnamefont{Cai}}, \bibinfo {author}
  {\bibfnamefont{X.}~\bibnamefont{Wu}}, \bibinfo {author}
  {\bibfnamefont{W.}~\bibnamefont{Ho}}, \bibinfo {author}
  {\bibfnamefont{C.}~\bibnamefont{Gao}}, \bibinfo {author}
  {\bibfnamefont{J.}~\bibnamefont{Jia}}, \bibinfo {author}
  {\bibfnamefont{N.}~\bibnamefont{Wang}}, \bibinfo {author}
  {\bibfnamefont{H.}~\bibnamefont{Fan}} \emph{et~al.},\ }%
  \Doi{10.1103/PhysRevLett.113.066105}{\emph{\bibinfo {title} {Dense Network of
  One-Dimensional Midgap Metallic Modes in Monolayer ${\mathrm{MoSe}}_{2}$ and
  Their Spatial Undulations}}},\ \bibfield{journal}{%
  \bibinfo {journal} {Phys. Rev. Lett.}\ }%
  \textbf{\bibinfo {volume} {113}},\ \bibinfo {pages} {066105} (\bibinfo {year}
  {2014}).%
  \bibAnnoteFile{Stop}{Liu2014}%
\bibitem{Barja2016}%
  \BibitemOpen
  \bibfield{author}{%
  \bibinfo {author} {\bibfnamefont{S.}~\bibnamefont{Barja}}, \bibinfo {author}
  {\bibfnamefont{S.}~\bibnamefont{Wickenburg}}, \bibinfo {author}
  {\bibfnamefont{Z.-F.}\ \bibnamefont{Liu}}, \bibinfo {author}
  {\bibfnamefont{Y.}~\bibnamefont{Zhang}}, \bibinfo {author}
  {\bibfnamefont{H.}~\bibnamefont{Ryu}}, \bibinfo {author}
  {\bibfnamefont{M.~M.}\ \bibnamefont{Ugeda}}, \bibinfo {author}
  {\bibfnamefont{Z.}~\bibnamefont{Hussain}}, \bibinfo {author}
  {\bibfnamefont{Z.-X.}\ \bibnamefont{Shen}}, \bibinfo {author}
  {\bibfnamefont{S.-K.}\ \bibnamefont{Mo}}, \bibinfo {author}
  {\bibfnamefont{E.}~\bibnamefont{Wong}} \emph{et~al.},\ }%
  \Doi{10.1038/nphys3730}{\emph{\bibinfo {title} {Charge density wave order in
  1D mirror twin boundaries of single-layer MoSe$_2$}}},\ \bibfield{journal}{%
  \bibinfo {journal} {Nat. Phys.}\ }%
  \textbf{\bibinfo {volume} {12}},\ \bibinfo {pages} {751} (\bibinfo {year}
  {2016}).%
  \bibAnnoteFile{Stop}{Barja2016}%
\bibitem{Peierls}%
  \BibitemOpen
  \bibfield{author}{%
  \bibinfo {author} {\bibfnamefont{R.}~\bibnamefont{Peierls}},\ }%
  \emph{\bibinfo {title} {Quantum Theory of Solids}}\ (\bibinfo {publisher}
  {Oxford Univ. Press},\ \bibinfo {year} {1955}).%
  \bibAnnoteFile{Stop}{Peierls}%
\bibitem{Lee385}%
  \BibitemOpen
  \bibfield{author}{%
  \bibinfo {author} {\bibfnamefont{C.}~\bibnamefont{Lee}}, \bibinfo {author}
  {\bibfnamefont{X.}~\bibnamefont{Wei}}, \bibinfo {author}
  {\bibfnamefont{J.~W.}\ \bibnamefont{Kysar}},\ and\ \bibinfo {author}
  {\bibfnamefont{J.}~\bibnamefont{Hone}},\ }%
  \Doi{10.1126/science.1157996}{\emph{\bibinfo {title} {Measurement of the
  Elastic Properties and Intrinsic Strength of Monolayer Graphene}}},\
  \bibfield{journal}{%
  \bibinfo {journal} {Science}\ }%
  \textbf{\bibinfo {volume} {321}},\ \bibinfo {pages} {385} (\bibinfo {year}
  {2008}).%
  \bibAnnoteFile{Stop}{Lee385}%
\bibitem{Akinwande2017}%
  \BibitemOpen
  \bibfield{author}{%
  \bibinfo {author} {\bibfnamefont{D.}~\bibnamefont{Akinwande}}, \bibinfo
  {author} {\bibfnamefont{C.~J.}\ \bibnamefont{Brennan}}, \bibinfo {author}
  {\bibfnamefont{J.~S.}\ \bibnamefont{Bunch}}, \bibinfo {author}
  {\bibfnamefont{P.}~\bibnamefont{Egberts}}, \bibinfo {author}
  {\bibfnamefont{J.~R.}\ \bibnamefont{Felts}}, \bibinfo {author}
  {\bibfnamefont{H.}~\bibnamefont{Gao}}, \bibinfo {author}
  {\bibfnamefont{R.}~\bibnamefont{Huang}}, \bibinfo {author}
  {\bibfnamefont{J.-S.}\ \bibnamefont{Kim}}, \bibinfo {author}
  {\bibfnamefont{T.}~\bibnamefont{Li}}, \bibinfo {author}
  {\bibfnamefont{Y.}~\bibnamefont{Li}} \emph{et~al.},\ }%
  \Doi{http://dx.doi.org/10.1016/j.eml.2017.01.008}{\emph{\bibinfo {title} {A
  review on mechanics and mechanical properties of 2D materials---Graphene and
  beyond}}},\ \bibfield{journal}{%
  \bibinfo {journal} {Extreme Mechanics Letters}\ }%
  \textbf{\bibinfo {volume} {13}},\ \bibinfo {pages} {42 } (\bibinfo {year}
  {2017}).%
  \bibAnnoteFile{Stop}{Akinwande2017}%
\bibitem{Duerloo2012}%
  \BibitemOpen
  \bibfield{author}{%
  \bibinfo {author} {\bibfnamefont{K.-A.~N.}\ \bibnamefont{Duerloo}}, \bibinfo
  {author} {\bibfnamefont{M.~T.}\ \bibnamefont{Ong}},\ and\ \bibinfo {author}
  {\bibfnamefont{E.~J.}\ \bibnamefont{Reed}},\ }%
  \Doi{10.1021/jz3012436}{\emph{\bibinfo {title} {Intrinsic Piezoelectricity in
  Two-Dimensional Materials}}},\ \bibfield{journal}{%
  \bibinfo {journal} {The Journal of Physical Chemistry Letters}\ }%
  \textbf{\bibinfo {volume} {3}},\ \bibinfo {pages} {2871} (\bibinfo {year}
  {2012}).%
  \bibAnnoteFile{Stop}{Duerloo2012}%
\bibitem{fei2015giant}%
  \BibitemOpen
  \bibfield{author}{%
  \bibinfo {author} {\bibfnamefont{R.}~\bibnamefont{Fei}}, \bibinfo {author}
  {\bibfnamefont{W.}~\bibnamefont{Li}}, \bibinfo {author}
  {\bibfnamefont{J.}~\bibnamefont{Li}},\ and\ \bibinfo {author}
  {\bibfnamefont{L.}~\bibnamefont{Yang}},\ }%
  \Doi{10.1063/1.4934750}{\emph{\bibinfo {title} {Giant piezoelectricity of
  monolayer group IV monochalcogenides: SnSe, SnS, GeSe, and GeS}}},\
  \bibfield{journal}{%
  \bibinfo {journal} {Applied Physics Letters}\ }%
  \textbf{\bibinfo {volume} {107}},\ \bibinfo {pages} {173104} (\bibinfo {year}
  {2015}).%
  \bibAnnoteFile{Stop}{fei2015giant}%
\bibitem{Gomes2015}%
  \BibitemOpen
  \bibfield{author}{%
  \bibinfo {author} {\bibfnamefont{L.~C.}\ \bibnamefont{Gomes}}, \bibinfo
  {author} {\bibfnamefont{A.}~\bibnamefont{Carvalho}},\ and\ \bibinfo {author}
  {\bibfnamefont{A.~H.}\ \bibnamefont{Castro~Neto}},\ }%
  \Doi{10.1103/PhysRevB.92.214103}{\emph{\bibinfo {title} {Enhanced
  piezoelectricity and modified dielectric screening of two-dimensional
  group-IV monochalcogenides}}},\ \bibfield{journal}{%
  \bibinfo {journal} {Phys. Rev. B}\ }%
  \textbf{\bibinfo {volume} {92}},\ \bibinfo {pages} {214103} (\bibinfo {year}
  {2015}).%
  \bibAnnoteFile{Stop}{Gomes2015}%
\bibitem{Blonsky2015}%
  \BibitemOpen
  \bibfield{author}{%
  \bibinfo {author} {\bibfnamefont{M.~N.}\ \bibnamefont{Blonsky}}, \bibinfo
  {author} {\bibfnamefont{H.~L.}\ \bibnamefont{Zhuang}}, \bibinfo {author}
  {\bibfnamefont{A.~K.}\ \bibnamefont{Singh}},\ and\ \bibinfo {author}
  {\bibfnamefont{R.~G.}\ \bibnamefont{Hennig}},\ }%
  \Doi{10.1021/acsnano.5b03394}{\emph{\bibinfo {title} {Ab Initio Prediction of
  Piezoelectricity in Two-Dimensional Materials}}},\ \bibfield{journal}{%
  \bibinfo {journal} {ACS Nano}\ }%
  \textbf{\bibinfo {volume} {9}},\ \bibinfo {pages} {9885} (\bibinfo {year}
  {2015}).%
  \bibAnnoteFile{Stop}{Blonsky2015}%
\bibitem{Fei2016}%
  \BibitemOpen
  \bibfield{author}{%
  \bibinfo {author} {\bibfnamefont{R.}~\bibnamefont{Fei}}, \bibinfo {author}
  {\bibfnamefont{W.}~\bibnamefont{Kang}},\ and\ \bibinfo {author}
  {\bibfnamefont{L.}~\bibnamefont{Yang}},\ }%
  \Doi{10.1103/PhysRevLett.117.097601}{\emph{\bibinfo {title} {Ferroelectricity
  and Phase Transitions in Monolayer Group-IV Monochalcogenides}}},\
  \bibfield{journal}{%
  \bibinfo {journal} {Phys. Rev. Lett.}\ }%
  \textbf{\bibinfo {volume} {117}},\ \bibinfo {pages} {097601} (\bibinfo {year}
  {2016}).%
  \bibAnnoteFile{Stop}{Fei2016}%
\bibitem{Hanakata2016}%
  \BibitemOpen
  \bibfield{author}{%
  \bibinfo {author} {\bibfnamefont{P.~Z.}\ \bibnamefont{Hanakata}}, \bibinfo
  {author} {\bibfnamefont{A.}~\bibnamefont{Carvalho}}, \bibinfo {author}
  {\bibfnamefont{D.~K.}\ \bibnamefont{Campbell}},\ and\ \bibinfo {author}
  {\bibfnamefont{H.~S.}\ \bibnamefont{Park}},\ }%
  \Doi{10.1103/PhysRevB.94.035304}{\emph{\bibinfo {title} {Polarization and
  valley switching in monolayer group-IV monochalcogenides}}},\
  \bibfield{journal}{%
  \bibinfo {journal} {Phys. Rev. B}\ }%
  \textbf{\bibinfo {volume} {94}},\ \bibinfo {pages} {035304} (\bibinfo {year}
  {2016}).%
  \bibAnnoteFile{Stop}{Hanakata2016}%
\bibitem{Chang2016}%
  \BibitemOpen
  \bibfield{author}{%
  \bibinfo {author} {\bibfnamefont{K.}~\bibnamefont{Chang}}, \bibinfo {author}
  {\bibfnamefont{J.}~\bibnamefont{Liu}}, \bibinfo {author}
  {\bibfnamefont{H.}~\bibnamefont{Lin}}, \bibinfo {author}
  {\bibfnamefont{N.}~\bibnamefont{Wang}}, \bibinfo {author}
  {\bibfnamefont{K.}~\bibnamefont{Zhao}}, \bibinfo {author}
  {\bibfnamefont{A.}~\bibnamefont{Zhang}}, \bibinfo {author}
  {\bibfnamefont{F.}~\bibnamefont{Jin}}, \bibinfo {author}
  {\bibfnamefont{Y.}~\bibnamefont{Zhong}}, \bibinfo {author}
  {\bibfnamefont{X.}~\bibnamefont{Hu}}, \bibinfo {author}
  {\bibfnamefont{W.}~\bibnamefont{Duan}} \emph{et~al.},\ }%
  \Doi{10.1126/science.aad8609}{\emph{\bibinfo {title} {Discovery of robust
  in-plane ferroelectricity in atomic-thick SnTe}}},\ \bibfield{journal}{%
  \bibinfo {journal} {Science}\ }%
  \textbf{\bibinfo {volume} {353}},\ \bibinfo {pages} {274} (\bibinfo {year}
  {2016}).%
  \bibAnnoteFile{Stop}{Chang2016}%
\bibitem{Janotti2012}%
  \BibitemOpen
  \bibfield{author}{%
  \bibinfo {author} {\bibfnamefont{A.}~\bibnamefont{Janotti}}, \bibinfo
  {author} {\bibfnamefont{L.}~\bibnamefont{Bjaalie}}, \bibinfo {author}
  {\bibfnamefont{L.}~\bibnamefont{Gordon}},\ and\ \bibinfo {author}
  {\bibfnamefont{C.~G.}\ \bibnamefont{Van~de Walle}},\ }%
  \Doi{10.1103/PhysRevB.86.241108}{\emph{\bibinfo {title} {Controlling the
  density of the two-dimensional electron gas at the
  SrTiO${}_{3}$/LaAlO${}_{3}$ interface}}},\ \bibfield{journal}{%
  \bibinfo {journal} {Phys. Rev. B}\ }%
  \textbf{\bibinfo {volume} {86}},\ \bibinfo {pages} {241108} (\bibinfo {year}
  {2012}).%
  \bibAnnoteFile{Stop}{Janotti2012}%
\bibitem{Keldysh1979}%
  \BibitemOpen
  \bibfield{author}{%
  \bibinfo {author} {\bibfnamefont{L.~V.}\ \bibnamefont{Keldysh}},\ }%
  \href{http://adsabs.harvard.edu/abs/1979JETPL..29..658K}{\emph{\bibinfo
  {title} {{Coulomb interaction in thin semiconductor and semimetal films}}}},\
  \bibfield{journal}{%
  \bibinfo {journal} {Soviet Journal of Experimental and Theoretical Physics}\
  }%
  \textbf{\bibinfo {volume} {29}},\ \bibinfo {pages} {658} (\bibinfo {year}
  {1979}).%
  \bibAnnoteFile{Stop}{Keldysh1979}%
\bibitem{cudazzo2011dielectric}%
  \BibitemOpen
  \bibfield{author}{%
  \bibinfo {author} {\bibfnamefont{P.}~\bibnamefont{Cudazzo}}, \bibinfo
  {author} {\bibfnamefont{I.~V.}\ \bibnamefont{Tokatly}},\ and\ \bibinfo
  {author} {\bibfnamefont{A.}~\bibnamefont{Rubio}},\ }%
  \Doi{10.1103/PhysRevB.84.085406}{\emph{\bibinfo {title} {Dielectric screening
  in two-dimensional insulators: Implications for excitonic and impurity states
  in graphane}}},\ \bibfield{journal}{%
  \bibinfo {journal} {Physical Review B}\ }%
  \textbf{\bibinfo {volume} {84}},\ \bibinfo {pages} {085406} (\bibinfo {year}
  {2011}).%
  \bibAnnoteFile{Stop}{cudazzo2011dielectric}%
\bibitem{Qiu2016}%
  \BibitemOpen
  \bibfield{author}{%
  \bibinfo {author} {\bibfnamefont{D.~Y.}\ \bibnamefont{Qiu}}, \bibinfo
  {author} {\bibfnamefont{F.~H.}\ \bibnamefont{da~Jornada}},\ and\ \bibinfo
  {author} {\bibfnamefont{S.~G.}\ \bibnamefont{Louie}},\ }%
  \Doi{10.1103/PhysRevB.93.235435}{\emph{\bibinfo {title} {Screening and
  many-body effects in two-dimensional crystals: Monolayer
  ${\mathrm{MoS}}_{2}$}}},\ \bibfield{journal}{%
  \bibinfo {journal} {Phys. Rev. B}\ }%
  \textbf{\bibinfo {volume} {93}},\ \bibinfo {pages} {235435} (\bibinfo {year}
  {2016}).%
  \bibAnnoteFile{Stop}{Qiu2016}%
\bibitem{Gia2009}%
  \BibitemOpen
  \bibfield{author}{%
  \bibinfo {author} {\bibfnamefont{P.}~\bibnamefont{Giannozzi}}, \bibinfo
  {author} {\bibfnamefont{S.}~\bibnamefont{Baroni}}, \bibinfo {author}
  {\bibfnamefont{N.}~\bibnamefont{Bonini}}, \bibinfo {author}
  {\bibfnamefont{M.}~\bibnamefont{Calandra}}, \bibinfo {author}
  {\bibfnamefont{R.}~\bibnamefont{Car}}, \bibinfo {author}
  {\bibfnamefont{C.}~\bibnamefont{Cavazzoni}}, \bibinfo {author}
  {\bibfnamefont{D.}~\bibnamefont{Ceresoli}}, \bibinfo {author}
  {\bibfnamefont{G.~L.}\ \bibnamefont{Chiarotti}}, \bibinfo {author}
  {\bibfnamefont{M.}~\bibnamefont{Cococcioni}}, \bibinfo {author}
  {\bibfnamefont{I.}~\bibnamefont{Dabo}} \emph{et~al.},\ }%
  \href{http://iopscience.iop.org/0953-8984/21/39/395502}{\emph{\bibinfo
  {title} {QUANTUM ESPRESSO: a modular and open-source software project for
  quantum simulations of materials}}},\ \bibfield{journal}{%
  \bibinfo {journal} {J. Phys. Condens. Matter}\ }%
  \textbf{\bibinfo {volume} {21}},\ \bibinfo {pages} {395502} (\bibinfo {year}
  {2009}).%
  \bibAnnoteFile{Stop}{Gia2009}%
\bibitem{PBE}%
  \BibitemOpen
  \bibfield{author}{%
  \bibinfo {author} {\bibfnamefont{J.~P.}\ \bibnamefont{Perdew}}, \bibinfo
  {author} {\bibfnamefont{K.}~\bibnamefont{Burke}},\ and\ \bibinfo {author}
  {\bibfnamefont{M.}~\bibnamefont{Ernzerhof}},\ }%
  \Doi{10.1103/PhysRevLett.77.3865}{\emph{\bibinfo {title} {Generalized
  Gradient Approximation Made Simple}}},\ \bibfield{journal}{%
  \bibinfo {journal} {Phys. Rev. Lett.}\ }%
  \textbf{\bibinfo {volume} {77}},\ \bibinfo {pages} {3865} (\bibinfo {year}
  {1996}).%
  \bibAnnoteFile{Stop}{PBE}%
\bibitem{GBRV}%
  \BibitemOpen
  \bibfield{author}{%
  \bibinfo {author} {\bibfnamefont{K.~F.}\ \bibnamefont{Garrity}}, \bibinfo
  {author} {\bibfnamefont{J.~W.}\ \bibnamefont{Bennett}}, \bibinfo {author}
  {\bibfnamefont{K.~M.}\ \bibnamefont{Rabe}},\ and\ \bibinfo {author}
  {\bibfnamefont{D.}~\bibnamefont{Vanderbilt}},\ }%
  \Doi{http://dx.doi.org/10.1016/j.commatsci.2013.08.053}{\emph{\bibinfo
  {title} {{Pseudopotentials for high-throughput {DFT} calculations}}}},\
  \bibfield{journal}{%
  \bibinfo {journal} {Computational Materials Science}\ }%
  \textbf{\bibinfo {volume} {81}},\ \bibinfo {pages} {446 } (\bibinfo {year}
  {2014}).%
  \bibAnnoteFile{Stop}{GBRV}%
\bibitem{SSSP}%
  \BibitemOpen
  \bibfield{author}{%
  \bibinfo {author} {\bibfnamefont{I.~E.}\ \bibnamefont{Castelli}}
  \emph{et~al.},\ }%
  {\bibinfo {title} {Standard solid state pseudopotentials ({SSSP})},}\
  \href{http://www.materialscloud.org/sssp/}{http://www.materialscloud.org/sssp/}.%
  \bibAnnoteFile{Stop}{SSSP}%
\bibitem{Monk1976}%
  \BibitemOpen
  \bibfield{author}{%
  \bibinfo {author} {\bibfnamefont{H.~J.}\ \bibnamefont{Monkhorst}}\ and\
  \bibinfo {author} {\bibfnamefont{J.~D.}\ \bibnamefont{Pack}},\ }%
  \Doi{10.1103/PhysRevB.13.5188}{\emph{\bibinfo {title} {Special points for
  Brillouin-zone integrations}}},\ \bibfield{journal}{%
  \bibinfo {journal} {Phys. Rev. B}\ }%
  \textbf{\bibinfo {volume} {13}},\ \bibinfo {pages} {5188} (\bibinfo {year}
  {1976}).%
  \bibAnnoteFile{Stop}{Monk1976}%
\bibitem{marzari1999thermal}%
  \BibitemOpen
  \bibfield{author}{%
  \bibinfo {author} {\bibfnamefont{N.}~\bibnamefont{Marzari}}, \bibinfo
  {author} {\bibfnamefont{D.}~\bibnamefont{Vanderbilt}}, \bibinfo {author}
  {\bibfnamefont{A.}~\bibnamefont{De~Vita}},\ and\ \bibinfo {author}
  {\bibfnamefont{M.~C.}\ \bibnamefont{Payne}},\ }%
  \Doi{10.1103/PhysRevLett.82.3296}{\emph{\bibinfo {title} {Thermal Contraction
  and Disordering of the Al(110) Surface}}},\ \bibfield{journal}{%
  \bibinfo {journal} {Phys. Rev. Lett.}\ }%
  \textbf{\bibinfo {volume} {82}},\ \bibinfo {pages} {3296} (\bibinfo {year}
  {1999}).%
  \bibAnnoteFile{Stop}{marzari1999thermal}%
\bibitem{Otani2006}%
  \BibitemOpen
  \bibfield{author}{%
  \bibinfo {author} {\bibfnamefont{M.}~\bibnamefont{Otani}}\ and\ \bibinfo
  {author} {\bibfnamefont{O.}~\bibnamefont{Sugino}},\ }%
  \Doi{10.1103/PhysRevB.73.115407}{\emph{\bibinfo {title} {First-principles
  calculations of charged surfaces and interfaces: A plane-wave nonrepeated
  slab approach}}},\ \bibfield{journal}{%
  \bibinfo {journal} {Phys. Rev. B}\ }%
  \textbf{\bibinfo {volume} {73}},\ \bibinfo {pages} {115407} (\bibinfo {year}
  {2006}).%
  \bibAnnoteFile{Stop}{Otani2006}%
\bibitem{suppl}%
  \BibitemOpen
  \bibinfo {note} {See Supplemental Material at [URL].}%
  \bibAnnoteFile{Stop}{suppl}%
\bibitem{brown1990numerical}%
  \BibitemOpen
  \bibfield{author}{%
  \bibinfo {author} {\bibfnamefont{H.}~\bibnamefont{Brown}}, \bibinfo {author}
  {\bibfnamefont{A.}~\bibnamefont{Whittaker}},\ and\ \bibinfo {author}
  {\bibfnamefont{N.}~\bibnamefont{Rollins}},\ }%
  \Doi{10.1016/0038-1101(90)90197-M}{\emph{\bibinfo {title} {A numerical
  analysis of the resonant states of quantum well and superlattice devices}}},\
  \bibfield{journal}{%
  \bibinfo {journal} {Solid-State Electronics}\ }%
  \textbf{\bibinfo {volume} {33}},\ \bibinfo {pages} {333} (\bibinfo {year}
  {1990}).%
  \bibAnnoteFile{Stop}{brown1990numerical}%
\bibitem{tan1990self}%
  \BibitemOpen
  \bibfield{author}{%
  \bibinfo {author} {\bibfnamefont{I.-H.}\ \bibnamefont{Tan}}, \bibinfo
  {author} {\bibfnamefont{G.}~\bibnamefont{Snider}}, \bibinfo {author}
  {\bibfnamefont{L.}~\bibnamefont{Chang}},\ and\ \bibinfo {author}
  {\bibfnamefont{E.}~\bibnamefont{Hu}},\ }%
  \href{http://aip.scitation.org/doi/abs/10.1063/1.346245}{\emph{\bibinfo
  {title} {A self-consistent solution of Schr{\"o}dinger--Poisson equations
  using a nonuniform mesh}}},\ \bibfield{journal}{%
  \bibinfo {journal} {Journal of Applied Physics}\ }%
  \textbf{\bibinfo {volume} {68}},\ \bibinfo {pages} {4071} (\bibinfo {year}
  {1990}).%
  \bibAnnoteFile{Stop}{tan1990self}%
\bibitem{pulay1980convergence}%
  \BibitemOpen
  \bibfield{author}{%
  \bibinfo {author} {\bibfnamefont{P.}~\bibnamefont{Pulay}},\ }%
  \Doi{10.1016/0009-2614(80)80396-4}{\emph{\bibinfo {title} {Convergence
  acceleration of iterative sequences. The case of SCF iteration}}},\
  \bibfield{journal}{%
  \bibinfo {journal} {Chemical Physics Letters}\ }%
  \textbf{\bibinfo {volume} {73}},\ \bibinfo {pages} {393} (\bibinfo {year}
  {1980}).%
  \bibAnnoteFile{Stop}{pulay1980convergence}%
\bibitem{serra1991one}%
  \BibitemOpen
  \bibfield{author}{%
  \bibinfo {author} {\bibfnamefont{A.~C.}\ \bibnamefont{Serra}}\ and\ \bibinfo
  {author} {\bibfnamefont{H.~A.}\ \bibnamefont{Santos}},\ }%
  \Doi{10.1063/1.349389}{\emph{\bibinfo {title} {A one-dimensional,
  self-consistent numerical solution of Schr{\"o}dinger and Poisson
  equations}}},\ \bibfield{journal}{%
  \bibinfo {journal} {Journal of Applied Physics}\ }%
  \textbf{\bibinfo {volume} {70}},\ \bibinfo {pages} {2734} (\bibinfo {year}
  {1991}).%
  \bibAnnoteFile{Stop}{serra1991one}%
\bibitem{karner2007multi}%
  \BibitemOpen
  \bibfield{author}{%
  \bibinfo {author} {\bibfnamefont{M.}~\bibnamefont{Karner}}, \bibinfo {author}
  {\bibfnamefont{A.}~\bibnamefont{Gehring}}, \bibinfo {author}
  {\bibfnamefont{S.}~\bibnamefont{Holzer}}, \bibinfo {author}
  {\bibfnamefont{M.}~\bibnamefont{Pourfath}}, \bibinfo {author}
  {\bibfnamefont{M.}~\bibnamefont{Wagner}}, \bibinfo {author}
  {\bibfnamefont{W.}~\bibnamefont{Goes}}, \bibinfo {author}
  {\bibfnamefont{M.}~\bibnamefont{Vasicek}}, \bibinfo {author}
  {\bibfnamefont{O.}~\bibnamefont{Baumgartner}}, \bibinfo {author}
  {\bibfnamefont{C.}~\bibnamefont{Kernstock}}, \bibinfo {author}
  {\bibfnamefont{K.}~\bibnamefont{Schnass}} \emph{et~al.},\ }%
  \Doi{10.1007/s10825-006-0077-7}{\emph{\bibinfo {title} {A multi-purpose
  Schr{\"o}dinger-Poisson solver for TCAD applications}}},\
  \bibfield{journal}{%
  \bibinfo {journal} {Journal of Computational Electronics}\ }%
  \textbf{\bibinfo {volume} {6}},\ \bibinfo {pages} {179} (\bibinfo {year}
  {2007}).%
  \bibAnnoteFile{Stop}{karner2007multi}%
\bibitem{Busby2010}%
  \BibitemOpen
  \bibfield{author}{%
  \bibinfo {author} {\bibfnamefont{Y.}~\bibnamefont{Busby}}, \bibinfo {author}
  {\bibfnamefont{M.}~\bibnamefont{De~Seta}}, \bibinfo {author}
  {\bibfnamefont{G.}~\bibnamefont{Capellini}}, \bibinfo {author}
  {\bibfnamefont{F.}~\bibnamefont{Evangelisti}}, \bibinfo {author}
  {\bibfnamefont{M.}~\bibnamefont{Ortolani}}, \bibinfo {author}
  {\bibfnamefont{M.}~\bibnamefont{Virgilio}}, \bibinfo {author}
  {\bibfnamefont{G.}~\bibnamefont{Grosso}}, \bibinfo {author}
  {\bibfnamefont{G.}~\bibnamefont{Pizzi}}, \bibinfo {author}
  {\bibfnamefont{P.}~\bibnamefont{Calvani}}, \bibinfo {author}
  {\bibfnamefont{S.}~\bibnamefont{Lupi}} \emph{et~al.},\ }%
  \Doi{10.1103/PhysRevB.82.205317}{\emph{\bibinfo {title} {Near- and
  far-infrared absorption and electronic structure of Ge-SiGe multiple quantum
  wells}}},\ \bibfield{journal}{%
  \bibinfo {journal} {Phys. Rev. B}\ }%
  \textbf{\bibinfo {volume} {82}},\ \bibinfo {pages} {205317} (\bibinfo {year}
  {2010}).%
  \bibAnnoteFile{Stop}{Busby2010}%
\bibitem{schrpoisson}%
  \BibitemOpen
  \href{http://github.com/giovannipizzi/schrpoisson\_2dmaterials}{http://github.com/giovannipizzi/schrpoisson\_2dmaterials}.%
  \bibAnnoteFile{Stop}{schrpoisson}%
\bibitem{Sohier2016}%
  \BibitemOpen
\bibfield{author}{%
    }%
  \bibfield{author}{%
  \bibinfo {author} {\bibfnamefont{T.}~\bibnamefont{Sohier}}, \bibinfo {author}
  {\bibfnamefont{M.}~\bibnamefont{Calandra}},\ and\ \bibinfo {author}
  {\bibfnamefont{F.}~\bibnamefont{Mauri}},\ }%
  \Doi{10.1103/PhysRevB.94.085415}{\emph{\bibinfo {title} {Two-dimensional
  Fr\"ohlich interaction in transition-metal dichalcogenide monolayers:
  Theoretical modeling and first-principles calculations}}},\
  \bibfield{journal}{%
  \bibinfo {journal} {Phys. Rev. B}\ }%
  \textbf{\bibinfo {volume} {94}},\ \bibinfo {pages} {085415} (\bibinfo {year}
  {2016}).%
  \bibAnnoteFile{Stop}{Sohier2016}%
\bibitem{King1992}%
  \BibitemOpen
  \bibfield{author}{%
  \bibinfo {author} {\bibfnamefont{R.~D.}\ \bibnamefont{King-Smith}}\ and\
  \bibinfo {author} {\bibfnamefont{D.}~\bibnamefont{Vanderbilt}},\ }%
  \Doi{10.1103/PhysRevB.47.1651}{\emph{\bibinfo {title} {Theory of polarization
  of crystalline solids}}},\ \bibfield{journal}{%
  \bibinfo {journal} {Phys. Rev. B}\ }%
  \textbf{\bibinfo {volume} {47}},\ \bibinfo {pages} {1651} (\bibinfo {year}
  {1993}).%
  \bibAnnoteFile{Stop}{King1992}%
\bibitem{Resta2007}%
  \BibitemOpen
  \bibfield{author}{%
  \bibinfo {author} {\bibfnamefont{R.}~\bibnamefont{Resta}}\ and\ \bibinfo
  {author} {\bibfnamefont{D.}~\bibnamefont{Vanderbilt}},\ }%
  in\ \emph{\bibinfo {booktitle} {Topics in Applied Physics}},\ Vol.\ \bibinfo
  {volume} {105}\ (\bibinfo {publisher} {Springer Berlin Heidelberg},\ \bibinfo
  {year} {2007})\ pp.\ \bibinfo {pages} {31--68}.%
  \bibAnnoteFile{Stop}{Resta2007}%
\bibitem{Bal1988}%
  \BibitemOpen
  \bibfield{author}{%
  \bibinfo {author} {\bibfnamefont{A.}~\bibnamefont{Baldereschi}}, \bibinfo
  {author} {\bibfnamefont{S.}~\bibnamefont{Baroni}},\ and\ \bibinfo {author}
  {\bibfnamefont{R.}~\bibnamefont{Resta}},\ }%
  \Doi{10.1103/PhysRevLett.61.734}{\emph{\bibinfo {title} {Band Offsets in
  Lattice-Matched Heterojunctions: A Model and First-Principles Calculations
  for GaAs/AlAs}}},\ \bibfield{journal}{%
  \bibinfo {journal} {Phys. Rev. Lett.}\ }%
  \textbf{\bibinfo {volume} {61}},\ \bibinfo {pages} {734} (\bibinfo {year}
  {1988}).%
  \bibAnnoteFile{Stop}{Bal1988}%
\bibitem{Yudin2013}%
  \BibitemOpen
  \bibfield{author}{%
  \bibinfo {author} {\bibfnamefont{P.~V.}\ \bibnamefont{Yudin}}\ and\ \bibinfo
  {author} {\bibfnamefont{A.~K.}\ \bibnamefont{Tagantsev}},\ }%
  \href{http://stacks.iop.org/0957-4484/24/i=43/a=432001}{\emph{\bibinfo
  {title} {Fundamentals of flexoelectricity in solids}}},\ \bibfield{journal}{%
  \bibinfo {journal} {Nanotechnology}\ }%
  \textbf{\bibinfo {volume} {24}},\ \bibinfo {pages} {432001} (\bibinfo {year}
  {2013}).%
  \bibAnnoteFile{Stop}{Yudin2013}%
\bibitem{Naumov2009}%
  \BibitemOpen
  \bibfield{author}{%
  \bibinfo {author} {\bibfnamefont{I.}~\bibnamefont{Naumov}}, \bibinfo {author}
  {\bibfnamefont{A.~M.}\ \bibnamefont{Bratkovsky}},\ and\ \bibinfo {author}
  {\bibfnamefont{V.}~\bibnamefont{Ranjan}},\ }%
  \Doi{10.1103/PhysRevLett.102.217601}{\emph{\bibinfo {title} {Unusual
  Flexoelectric Effect in Two-Dimensional Noncentrosymmetric $s{p}^{2}$-Bonded
  Crystals}}},\ \bibfield{journal}{%
  \bibinfo {journal} {Phys. Rev. Lett.}\ }%
  \textbf{\bibinfo {volume} {102}},\ \bibinfo {pages} {217601} (\bibinfo {year}
  {2009}).%
  \bibAnnoteFile{Stop}{Naumov2009}%
\bibitem{Duerloo2014}%
  \BibitemOpen
  \bibfield{author}{%
  \bibinfo {author} {\bibfnamefont{K.-A.~N.}\ \bibnamefont{Duerloo}}, \bibinfo
  {author} {\bibfnamefont{Y.}~\bibnamefont{Li}},\ and\ \bibinfo {author}
  {\bibfnamefont{E.~J.}\ \bibnamefont{Reed}},\ }%
  \Doi{10.1038/ncomms5214}{\emph{\bibinfo {title} {Structural phase transitions
  in two-dimensional Mo- and W-dichalcogenide monolayers}}},\
  \bibfield{journal}{%
  \bibinfo {journal} {Nature Communications}\ }%
  \textbf{\bibinfo {volume} {5}},\ \bibinfo {pages} {4214} (\bibinfo {year}
  {2014}).%
  \bibAnnoteFile{Stop}{Duerloo2014}%
\end{thebibliography}
\end{document}